\documentclass[journal]{IEEEtran}
\usepackage{graphicx,amssymb,color,bm}
\usepackage{psfrag} 
\usepackage{multirow,array,graphicx,,xcolor,colortbl,hhline,setspace}
\definecolor{myblue}{RGB}{0,112,192}
\usepackage[colorlinks=true,linkcolor=black,anchorcolor=black,citecolor=black,urlcolor=cyan]{hyperref}

\usepackage{cite}

\ifCLASSINFOpdf
\else
\fi
\usepackage{amsmath}
\ifCLASSOPTIONcompsoc
  \usepackage[caption=false,font=normalsize,labelfont=sf,textfont=sf]{subfig}
\else
  \usepackage[caption=false,font=footnotesize]{subfig}
\fi
\begin{document}
%
\title{Convolutional Neural Networks \\ to Enhance Coded Speech}
%
%
%

\author{Ziyue~Zhao,~ 
		Huijun~Liu,~
        Tim~Fingscheidt,~\IEEEmembership{Senior~Member,~IEEE,}
}

\maketitle

\begin{abstract}
  
Enhancing coded speech suffering from far-end acoustic background noise, quantization noise, and potentially transmission errors, is a challenging task. In this work we propose two postprocessing approaches applying convolutional neural networks (CNNs) either in the time domain or the cepstral domain to enhance the coded speech without any modification of the codecs. The time domain approach follows an end-to-end fashion, while the cepstral domain approach uses analysis-synthesis with cepstral domain features. The proposed postprocessors in both domains are evaluated for various narrowband and wideband speech codecs in a wide range of conditions. The proposed postprocessor improves speech quality (PESQ) by  up to 0.25 MOS-LQO points for G.711, 0.30 points for G.726, 0.82 points for G.722, and 0.26 points for adaptive multirate wideband codec (AMR-WB). In a subjective CCR listening test, the proposed postprocessor on G.711-coded speech exceeds the speech quality of an ITU-T-standardized postfilter by 0.36 CMOS points, and obtains a clear preference of 1.77 CMOS points compared to legacy G.711, \emph{even better than uncoded speech with statistical significance}. The source code for the cepstral domain approach to enhance G.711-coded speech is made available\footnote{ \href{https://github.com/ifnspaml/Enhancement-Coded-Speech}{https://github.com/ifnspaml/Enhancement-Coded-Speech}}.

\end{abstract}


%
\IEEEpeerreviewmaketitle

\section{Introduction} \label{sec_intro}
%
%
%
%
\IEEEPARstart{S}{peech} signals being subject to speech encoding, transmission, and decoding are often called transcoded speech, or simply: coded speech. Coded speech often suffers from far-end acoustic background noise, quantization noise, and potentially transmission errors. To enhance the quality of coded speech, postprocessing methods, operating just after speech decoding can be advantageously employed. 

To combat quantization noise at the receiver, a postfilter based on classical Wiener theory of optimal filtering has been standardized for the logarithmic pulse code modulation (PCM) G.711 codec~\cite{ITUTG711}. It is part of the G.711 audio quality enhancement toolbox~\cite{ITUTG711tollbox}, described in detail in the appendix of G.711.1~\cite{ITUTG7111NEW}, a wideband extension of G.711. This postfilter uses \emph{a priori} information on the A- or $\mu$-law properties to estimate the quantization noise power spectral density (PSD), assuming the quantization noise to be spectrally white~\cite{garcia2008pcm,hiwasaki2008g}. Then, a Wiener filter is derived by the estimation of the \emph{a priori} signal-to-noise-ratio (SNR) based on a two-step noise reduction approach~\cite{plapous2004two}. After the filtering process, a limitation of distortions is performed to control the waveform difference between the original signal and the postprocessed coded signal.

However, as the bitrates go down for most of the modern codecs, it becomes more difficult for the classical Wiener filter to effectively suppress the quantization noise, while maintaining the speech perceptually undistorted, since the SNR drops and more importantly, only the mean squared error (MSE) is minimized in the Wiener filter~\cite{ramamoorthy1984enhancement}. Therefore, some perceptually-based postfilters have been proposed to reduce the perceptual degradation caused by low bitrate codecs. Formant enhancement postfilters~\cite{ramamoorthy1988enhancement,chen1995adaptive} emphasize the peaks of the spectral envelope while further suppressing the valleys to reduce the impact of quantization noise in coded speech, since the formants are perceptually more important than the spectral valleys. This type of postfilter typically consists of three parts~\cite{chen1995adaptive}: The core short-term postfilter to enhance the formants, a tilt correction filter to compensate the low-pass tilt caused by the core postfilter, and an adaptive gain control to compensate the gain misadjustment caused by parts one and two. 

In addition to modifying the spectral envelope of the speech signal, the spectral fine structure of voiced speech is improved by a pitch enhancement postfilter, aiming to emphasize the harmonic peaks and attenuate the gaps between the harmonics~\cite{chen1995adaptive}. In practice, this long-term postfilter is always applied to low frequencies, where harmonic peaks are more prominent, which actually forms a bass postfilter~\cite{backstrom2017speech}. This bass postfilter and the formant enhancement postfilter are used either together or separately in the decoders of some standard codecs, e.g., in adaptive multi-rate (AMR)~\cite{AMR3GPP}, wideband AMR (AMR-WB)~\cite{AMRWB3GPP} and enhanced voice services (EVS)~\cite{EVS3GPP}. 

For speech codecs using the so-called algebraic code-excited linear prediction (ACELP) codebooks, e.g., AMR and AMR-WB, an anti-sparseness postprocessing procedure is applied, aiming to suppress the perceptual artifacts caused by the sparseness of the algebraic fixed codebook vectors with only a few non-zero pulses per subframe, especially in low bitrate modes~\cite{AMR3GPP,AMRWB3GPP}. A modification of the fixed codebook vector is adaptively selected based on the quantized adaptive codebook gain~\cite{hagen1998removal}. 

In an attempt to combat quantization noise, it has been shown that if residual correlation exists in coded signals~\cite{han_LMQ_1D_sigma_EUSIPCO,han_LMQ_formula_SCC,zhao2016improving} or more specifically, coded speech~\cite{han_adpmcadaptive}, a time-variant receiver-sided codebook or a shallow neural network can provide some gains in a system-compatible fashion. 

Apart from the aforementioned quantization noise, also far-end acoustic background noise can degrade the quality and intelligibility of coded speech. In most cases, noise reduction approaches are conducted as a transmitter-sided preprocessing step to suppress the background noise before the speech signal is coded and transmitted~\cite{varyspeech}.  
However, since the noise usually cannot be entirely suppressed and therefore speech with some residual noise is coded and transmitted to the receiver side, one can aim to further reduce the noise of the coded speech in the postprocessing procedure. To accomplish this, a modified postfilter has been proposed for speech quality enhancement, where the parameters corresponding to the formant and pitch emphasis are adaptively updated based on the statistics of the background noise~\cite{grancharov2008generalized}. Furthermore, in adverse noise conditions, also postfiltering methods to improve the speech intelligibility have been studied~\cite{jokinen2014adaptive}. 
Additionally, a kind of postprocessing to enhance the coded speech in transmitter-sided noisy environments by restoring the distorted background noise while masking main coding artifacts for low bitrate speech coding is proposed and standardized in EVS as comfort noise addition~\cite{EVS3GPP}. An artificial comfort noise is generated and added to the coded speech signal after the level and the spectral shape of the background noise are estimated~\cite{fuchs2015comfort}.

Recently, speech enhancement based on neural networks has been intensively studied~\cite{wang2013towards,narayanan2013ideal,xu2014experimental,xu2015regression,lu2012speech,lu2013speech,maas2012recurrent,weninger2015speech,lee2017deep,park2016fully,fu2016snr,kounovsky2017single,fu2017complex,fu2017raw,fu2017end}. Deep neural networks (DNNs) are used as a classification method to estimate the ideal binary mask~\cite{wang2013towards} or smoothed ideal ratio mask~\cite{narayanan2013ideal} for noise reduction. Also, some regression approaches based on DNNs to learn a mapping function from noisy to clean speech features have been proposed~\cite{xu2014experimental,xu2015regression}. Furthermore, a deep denoising autoencoder is applied for noise reduction, with either both clean pairs~\cite{lu2012speech} or noisy and clean pairs~\cite{lu2013speech} as inputs and targets to train the autoencoder. Besides, recurrent neural networks (RNNs) are used for speech enhancement, e.g., a recurrent denoising autoencoder for robust automatic speech recognition (ASR)~\cite{maas2012recurrent} and long short-term memory (LSTM) structure for noise reduction~\cite{weninger2015speech,lee2017deep}.

In addition to the DNNs and RNNs, convolutional neural networks (CNNs) are achieving increasing attention for the speech enhancement task~\cite{park2016fully,fu2016snr,kounovsky2017single,fu2017complex,fu2017raw,fu2017end}. The CNNs are trained to learn a mapping between the noisy speech features and the clean speech features, e.g., log-power spectrum~\cite{park2016fully,fu2016snr,kounovsky2017single} or complex spectrogram~\cite{fu2017complex}, or a mapping directly between the noisy raw speech waveform and clean raw speech waveform~\cite{fu2017raw,fu2017end}. The convolutional layers in the CNNs have the property of local filtering, i.e., the input features share the same network weights, resulting in translation invariance for the output of the network, which is a desired property for the modeling of speech~\cite{abdel2012applying}. This local filtering property makes the CNNs have the ability to characterize local information of the speech signal, which clearly provides benefits for the task of speech enhancement. It is also because of this property that the number of the trainable weights is reduced in a large scale compared to DNNs and RNNs with fully-connected structures, making it more efficient to train the network~\cite{park2016fully}. 

In this work, we use CNNs to enhance \emph{coded} speech, so that this operation can be seen as a postprocessor after speech decoding (or anywhere later in the transmission chain) aiming at improving speech quality at the far-end, which is different to the aforementioned noise reduction approaches. Fig.\ \ref{postproc_main} shows the general flow chart of postprocessing for coded speech.
Motivated by the successful application of CNNs to the image super-resolution problem in computer vision~\cite{dong2016image,mao2016image,shi2016real,kim2016accurate}, aiming at restoring the missing information from the low-resolution image, we propose to use similar convolutional network structures to restore improved speech from speech being subject to encoding and decoding. In terms of the topology, we adopt the deep convolutional encoder-decoder network topology~\cite{mao2016image}, which is a symmetric structure with multiple layers of convolution and deconvolution~\cite{noh2015learning,shelhamer2017fully}, in order to firstly preserve the major information of the input features and meanwhile reduce the corruption and then recover the details of the features~\cite{mao2016image,shi2016real}. Furthermore, skip-layer connections are added symmetrically between the convolution and deconvolution layers to form a residual network for an effective training~\cite{kim2016accurate,he2016deep}. 

The contribution of this work is threefold: First, based on the CNN topology, we propose two different postprocessing approaches in the time domain and the cepstral domain to restore the speech either in an end-to-end fashion or in an analysis-synthesis fashion with cepstral domain features. To our knowledge, it is the first time that deep learning methods are used to enhance coded speech. Second, we show by objective and subjective listening quality assessment that both proposed approaches show superior performance compared to the state of the art G.711 postprocessing. Finally, both proposed approaches are system-compatible for different kinds of codecs without any modification of the encoder or decoder. The simulation results in clean and noisy speech conditions, tandeming, and frame loss conditions show their effectiveness for some widely used speech codecs in narrowband and wideband.

The article is structured as follows: In Section \ref{sec_state_of_art} we briefly sketch state of the art G.711 postprocessing, which serves as a baseline method in the evaluation part. Next, we describe the proposed CNN postprocessing approaches in both time domain and cepstral domain in Section \ref{sec_cnn_post_proc}. Subsequently, the experimental setup and the instrumental metrics for speech quality evaluation are explained in Section \ref{sec_exp_set_metric}. Then, in Section \ref{sec_exp_eva_dis}, we present the evaluation results and discussion. Finally, we conclude our work in Section \ref{sec_conclusion}. 

\begin{psfrags}
	\begin{figure}[tp]
		
		\psfrag{a}[cc][cc]{\footnotesize $\tilde{s}(n)$}
		\psfrag{d}[cc][cc]{\footnotesize ${s}(n)$}
		\psfrag{f}[cc][cc]{\footnotesize $\hat{s}(n)$}
		
		\psfrag{b}[cc][cc]{\footnotesize Source}
		\psfrag{k}[ct][ct]{\footnotesize Speech}
		\psfrag{c}[cc][cc]{\footnotesize Bitstream}
		\psfrag{e}[cc][cc]{\footnotesize Coded}
		\psfrag{s}[ct][ct]{\footnotesize Speech}
		\psfrag{g}[cc][cr]{\footnotesize Enhanced}
		\psfrag{m}[cc][br]{\footnotesize Speech}
		
		\psfrag{h}[cc][cc]{\footnotesize Speech}
		\psfrag{n}[cc][cc]{\footnotesize Encoder}
		\psfrag{i}[cc][cc]{\footnotesize Speech}
		\psfrag{r}[cc][cc]{\footnotesize Decoder}
		\psfrag{j}[cc][cc]{\footnotesize Post-}
		\psfrag{q}[ct][ct]{\footnotesize processor}
	
		\centering
		\includegraphics[width=0.49\textwidth]{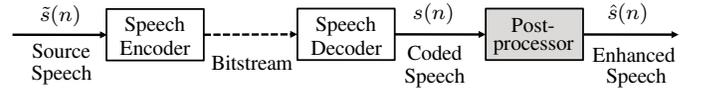}
		\vspace{-3pt}
		\caption{General flow chart of postprocessing for enhancement of coded speech.}
		\label{postproc_main}
	\end{figure}

\section{The G.711 Postprocessing Baseline} \label{sec_state_of_art}
In Fig. \ref{g711pf} the G.711 postprocessing aiming at attenuation of quantization noise is depicted. It has originally been proposed in~\cite{garcia2008pcm} and standardized in~\cite{ITUTG711tollbox}, basically following the classical framework of noise reduction, comprising: quantization noise power spectral density (PSD) estimation, \emph{a priori} SNR estimation, spectral weighting rule using the Wiener filter, and finally a quantization constraint. In the following subsections, this G.711 postprocessing is briefly reviewed as our baseline for enhancement of G.711-coded speech.

\vspace{-2pt}
\subsection{Quantization Noise PSD Estimation}
At first the coded speech $s(n)$ is subject to a periodic Hann window and then being transformed to the frequency domain representation $S(\ell,k)$ via the fast Fourier transform (FFT), with $\ell$ being the frame index and $k$ being the frequency bin index.
Since the quantization noise of G.711 is assumed to be spectrally white, the estimate of the quantization noise variance $\sigma_n^2(\ell)$ is sufficient for the quantization noise PSD estimation. To achieve this, an estimate of the (uncoded) source speech signal variance $\hat{\sigma}_{\tilde{s}}^2(\ell)$ is needed first, and subsequently an estimate of the load factor, defined as $\hat{\Gamma}(\ell)=1/\hat{\sigma}_{\tilde{s}}(\ell)$ denoting how the signal exploits the quantizer dynamic, is achieved. Interestingly, the estimate of the uncoded signal variance $\hat{\sigma}_{\tilde{s}}^2(\ell)$ is actually obtained by estimating the coded signal variance $\hat{\sigma}_s^2(\ell)$, assuming the variance of the quantization noise to be very low compared to the uncoded signal most of the time~\cite{garcia2008pcm}: 
\vspace{-5pt}
\begin{equation}
\hat{\sigma}_{\tilde{s}}^2(\ell) \approx \hat{\sigma}_s^2(\ell)=\frac{1}{\left | \mathcal{N_{\ell}} \right | }\sum_{n\in \mathcal{N_{\ell}}}s^2(n).
\end{equation}
The set $\mathcal{N_{\ell}}$ contains all sample indices $n$ belonging to frame $\ell$ and $\left | \mathcal{N_{\ell}} \right |$ is the number of samples in the frame. Then the signal-to-quantization-noise ratio is obtained according to the estimated load factor $\hat{\Gamma}(\ell)$ and the A- or $\mu$-law function. Finally, the estimate of the (spectrally white) quantization noise variance $\hat{\sigma}_n^2(\ell)$ is obtained. 

\begin{figure}[tp]
	
	\psfrag{a}[cc][cc]{\footnotesize ${s}(n)$}
	\psfrag{b}[cc][cc]{\footnotesize ${S}(\ell,k)$}
	\psfrag{c}[cc][cc]{\footnotesize $\hat{\xi}_2(\ell,k)$}
	\psfrag{d}[cl][cl]{\footnotesize $G_2(\ell,k)$}
	\psfrag{e}[cl][cl]{\footnotesize $\hat{\sigma}_n^2(\ell)$}
	\psfrag{f}[cl][cl]{\footnotesize $g_2(n)$}
	\psfrag{g}[cc][cc]{\footnotesize $\hat{s}(n)$}
	
	\psfrag{m}[cc][cc]{\footnotesize Windowing}
	\psfrag{2}[cc][cc]{\footnotesize and FFT}
	
	\psfrag{3}[cc][cc]{\footnotesize \emph{A priori} SNR }
	\psfrag{4}[cc][cc]{\footnotesize Estimation}
	
	\psfrag{5}[cc][cc]{\footnotesize Spectral Gain }
	\psfrag{6}[bc][br]{\footnotesize Function}
	
	\psfrag{7}[cc][cc]{\footnotesize IFFT and}
	\psfrag{8}[cc][cc]{\footnotesize Windowing}
	
	\psfrag{9}[cc][cc]{\footnotesize Quantization}
	\psfrag{0}[cc][cc]{\footnotesize Noise PSD}
	\psfrag{h}[cc][cc]{\footnotesize Estimation}
	
	\psfrag{i}[cc][cc]{\footnotesize Filtering}
	\psfrag{j}[cc][cc]{\footnotesize and OLS}
	
	\psfrag{k}[cc][cc]{\footnotesize Quantization}
	\psfrag{l}[cc][cc]{\footnotesize Constraint}
	
	\psfrag{y}[cl][cl]{\footnotesize $\hat{s}_2(n)$}
	
	\centering
	\includegraphics[width=0.49\textwidth]{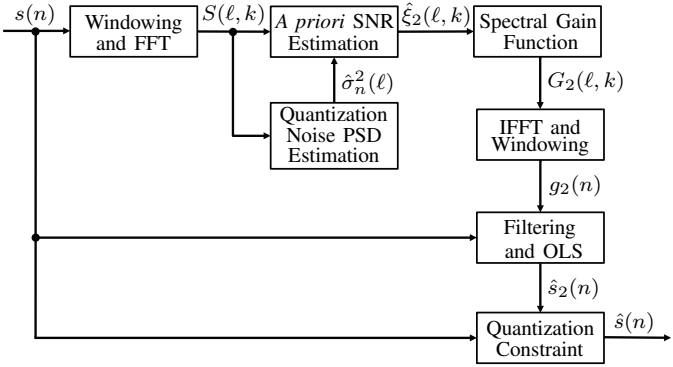}
	\vspace{-4pt}
	\caption{The postprocessing flow chart of the G.711 Amendment 2: New Appendix \uppercase\expandafter{\romannumeral3} audio quality enhancement toolbox (see~\cite{ITUTG711tollbox}).}
	\label{g711pf}
\end{figure}

\vspace{-5pt}
\subsection{A priori SNR Estimation and Wiener Filtering}
After estimation of the noise PSD, the \emph{a priori} SNR is obtained by a two-step noise reduction technique~\cite{plapous2004two} and subsequently the Wiener filter results. In order to estimate the \emph{a priori} SNR, the \emph{a posteriori} SNR is computed first as
\begin{equation}
\gamma (\ell,k)=\frac{\left | S(\ell,k) \right |^2}{\hat{\sigma}_n^2(\ell) }.
\end{equation}
Then, the first-step spectral gain function $G_1(\ell,k)$ from the Wiener filter can be expressed as 
\begin{equation}
G_1(\ell,k)=\frac{\hat{\xi}_1(\ell,k)}{1+\hat{\xi}_1(\ell,k)},
\end{equation}
where the first-step \emph{a priori} SNR estimate $\hat{\xi}_1(\ell,k)$ from the decision-direction approach~\cite{ephraim1984speech} is
\vspace{-3pt}
\begin{equation}
\hat{\xi}_1(\ell,k)\!=\!\beta\frac{| \hat{S}_1(\ell\!-\!1,k)  |^2}{\hat{\sigma}_n^2(\ell\!-\!1)}+(1\!-\!\beta)\text{max} \big(  \gamma (\ell,k)\!-\!1,0  \big ),  
\end{equation}
with $\hat{S}_1(\ell\!-\!1,k)\!=\!G_1(\ell\!-\!1,k)S(\ell\!-\!1,k)$ and $\beta$ being a weighting factor.
In the second step, an updated spectral gain function is computed as
\vspace{-5pt}
\begin{equation}
\label{equ_g2}
 G_2(\ell,k)=\text{max}\left ( \frac{\hat{\xi}_2(\ell,k)}{1+\hat{\xi}_2(\ell,k)},G_{\text{min}} \right ),
\end{equation}
where $G_{\text{min}}$ is the lower limit to avoid over-attenuation and
\begin{equation}
\hat{\xi}_2(\ell,k)=\frac{\left | G_1(\ell,k)S(\ell,k) \right |^2}{\hat{\sigma}_n^2(\ell)}
\end{equation} 
is the updated \emph{a priori} SNR estimate.
Finally, a causal filter impulse response $g_2(n)$ is obtained from this updated spectral gain function (\ref{equ_g2}) by inverse FFT (IFFT) and imposing a linear phase, and the coded speech $s(n)$ is time-domain-filtered and the overlap and save (OLS) method provides the enhanced speech $\hat{s}_2(n)$. Note that due to its frame structure, the G.711 postfilter baseline has an algorithmic delay of 2 ms.

\vspace{-8pt}
\subsection{Quantization Constraint}
\label{quant_cons}
\vspace{-1pt}
In order to avoid extra distortion introduced by the above postprocessing, finally a limitation of potential distortions is performed. Since the quantization interval of each coded speech sample $s(n)$ is known, this idea is to limit the postprocessed samples $\hat{s}(n)$ to lie within the respective interval. If an outlier sample (outside the certain quantization interval) is detected, the constraint will replace it by the closest decision boundary of this respective quantization interval. After application of this constraint, the final postprocessed speech $\hat{s}(n)$ is obtained.

\vspace{-3pt}
\section{Convolutional Neural Network (CNN) Postprocessing} \label{sec_cnn_post_proc}
\vspace{-1pt}
In this section, we present the proposed CNN-based postprocessing for coded speech alternatively in the time domain and in the cepstral domain. Fig. 3 depicts the high-level block diagram. \textcolor{black}{
	At first, for both approaches, the coded speech $s(n)$ is assembled to frames $\mathbf{s}(\ell)$, applying a window function. Then, the frame is processed either in the time domain resulting in $\hat{\mathbf{s}}_{\text{t}}(\ell)$, or in the cepstral domain resulting in $\hat{\mathbf{s}}_{\text{c}}(\ell)$. Finally, the enhanced speech $\hat{s}(n)$ is obtained via either a direct concatenation of the processed frames $\hat{\mathbf{s}}_{\text{t}}(\ell)$ for the time domain approach, or some waveform reconstruction of the processed frames $\hat{\mathbf{s}}_{\text{c}}(\ell)$ for the cepstral domain approach, as outlined in the following.
}

\begin{figure}[!t]
	
	\psfrag{a}[cc][cc]{\footnotesize ${s}(n)$}
	\psfrag{b}[cc][cc]{\footnotesize Windowing}
	\psfrag{c}[cc][cc]{\footnotesize $\mathbf{s}(\ell)$}
	
	\psfrag{z}[cc][cc]{\footnotesize ${s}(n)$}
	\psfrag{y}[cc][cc]{\footnotesize Windowing}
	\psfrag{o}[cc][cc]{\footnotesize $\mathbf{s}(\ell)$}

	\psfrag{e}[bc][bc]{\footnotesize Time Domain}
	\psfrag{d}[tc][tc]{\footnotesize Processing}
	
	\psfrag{i}[cc][cc]{\footnotesize Cepstral Domain}
	\psfrag{j}[tc][tc]{\footnotesize Processing}
	
	\psfrag{g}[cc][cr]{\footnotesize $\hat{\mathbf{s}}_{\text{t}}(\ell)$}
	\psfrag{k}[cc][cr]{\footnotesize $\hat{\mathbf{s}}_{\text{c}}(\ell)$}
	
	\psfrag{u}[bc][bc]{\footnotesize Concatenation}
	\psfrag{v}[cc][cr]{\footnotesize $\hat{s}(n)$}

	\psfrag{l}[cc][cc]{\footnotesize Waveform}
	\psfrag{n}[cc][cc]{\footnotesize Reconstruction}
	\psfrag{m}[cc][cr]{\footnotesize $\hat{s}(n)$}

	\centering
	\includegraphics[width=0.49\textwidth]{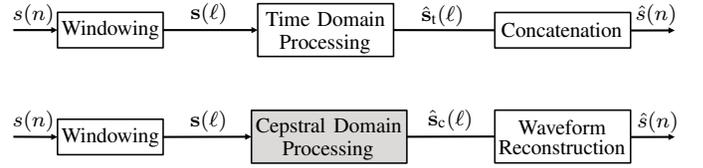}
	\vspace{-15pt}
	\textcolor{black}{ 
	\caption{{\bf CNN-based postprocessing} for {\bf time domain} approach (upper) and {\bf cepstral domain} approach (lower). More details of the cepstral domain processing can be found in Figs.~\ref{figures_postprocess_frame}, \ref{figures_postprocess_cepstral_dom}, and \ref{figures_CNN_topology}.}
	}
	\label{figures_postprocess_main}
\end{figure}

\begin{figure*}[!t]
	\psfrag{h}[cr][cr]{\footnotesize Windowing}
	
	\psfrag{2}[cr][cl]{\footnotesize Processing}
	\psfrag{3}[cr][br]{\footnotesize Approaches}
	
	\psfrag{i}[cr][cr]{\footnotesize Waveform}
	\psfrag{c}[cr][cr]{\footnotesize Reconstruction}

	\psfrag{a}[tc][tc]{\footnotesize $\hat{s}(n)$}	
	\psfrag{j}[cr][cr]{\footnotesize $s(n)$}
	
	\psfrag{0}[cc][cc]{\tiny $0\cdots 0$}
	\psfrag{y}[cc][cc]{\tiny $0 ... 0$}
	\psfrag{z}[cc][cc]{\tiny $0 ... 0$}
	\psfrag{t}[cc][cl]{\tiny $0 ... 0$}
	\psfrag{u}[cc][cc]{\tiny $\mathbf{0}$}
	\psfrag{v}[cc][cl]{\tiny $0 ... 0$}
	
	\psfrag{b}[cc][cc]{\footnotesize \textcolor{white}{Processing}}
	\psfrag{e}[cc][cc]{\footnotesize \textcolor{white}{Drop\,Past}}
	\psfrag{f}[tc][tl]{\footnotesize \textcolor{white}{OLA}}
	\psfrag{g}[cc][cc]{\footnotesize \textcolor{white}{OLA}}
	\psfrag{n}[cc][cc]{\footnotesize \textcolor{white}{Drop\,Past}}
	\psfrag{o}[cc][cc]{\footnotesize \textcolor{white}{OLA}}
	\psfrag{p}[cc][cc]{\footnotesize \textcolor{white}{OLA}}
	
	\psfrag{w}[cc][cl]{\footnotesize \uppercase\expandafter{\romannumeral1}}
	\psfrag{x}[cc][cl]{\footnotesize \uppercase\expandafter{\romannumeral2}}
	\psfrag{d}[cc][cl]{\footnotesize \uppercase\expandafter{\romannumeral3}}
	\psfrag{q}[cc][cc]{\footnotesize \uppercase\expandafter{\romannumeral4}}
	\psfrag{r}[cc][cc]{\footnotesize \uppercase\expandafter{\romannumeral5}}
	\psfrag{s}[cc][cc]{\footnotesize \uppercase\expandafter{\romannumeral6}}
	
	\psfrag{4}[bl][bl]{\footnotesize $10$ ms}
	\psfrag{5}[bl][bl]{\footnotesize $10$ ms}
	\psfrag{6}[bl][bl]{\footnotesize $10$ ms}
	\psfrag{7}[bl][bl]{\footnotesize $20$ ms}
	\psfrag{8}[bl][bl]{\footnotesize $20$ ms}
	\psfrag{9}[bl][bl]{\footnotesize $16$ ms}
	
	\psfrag{A}[cr][cr]{\footnotesize Framework}
	\psfrag{B}[cr][br]{\footnotesize Structure Num.}
	
	\centering
	\subfloat{\includegraphics[width=0.5\textwidth]{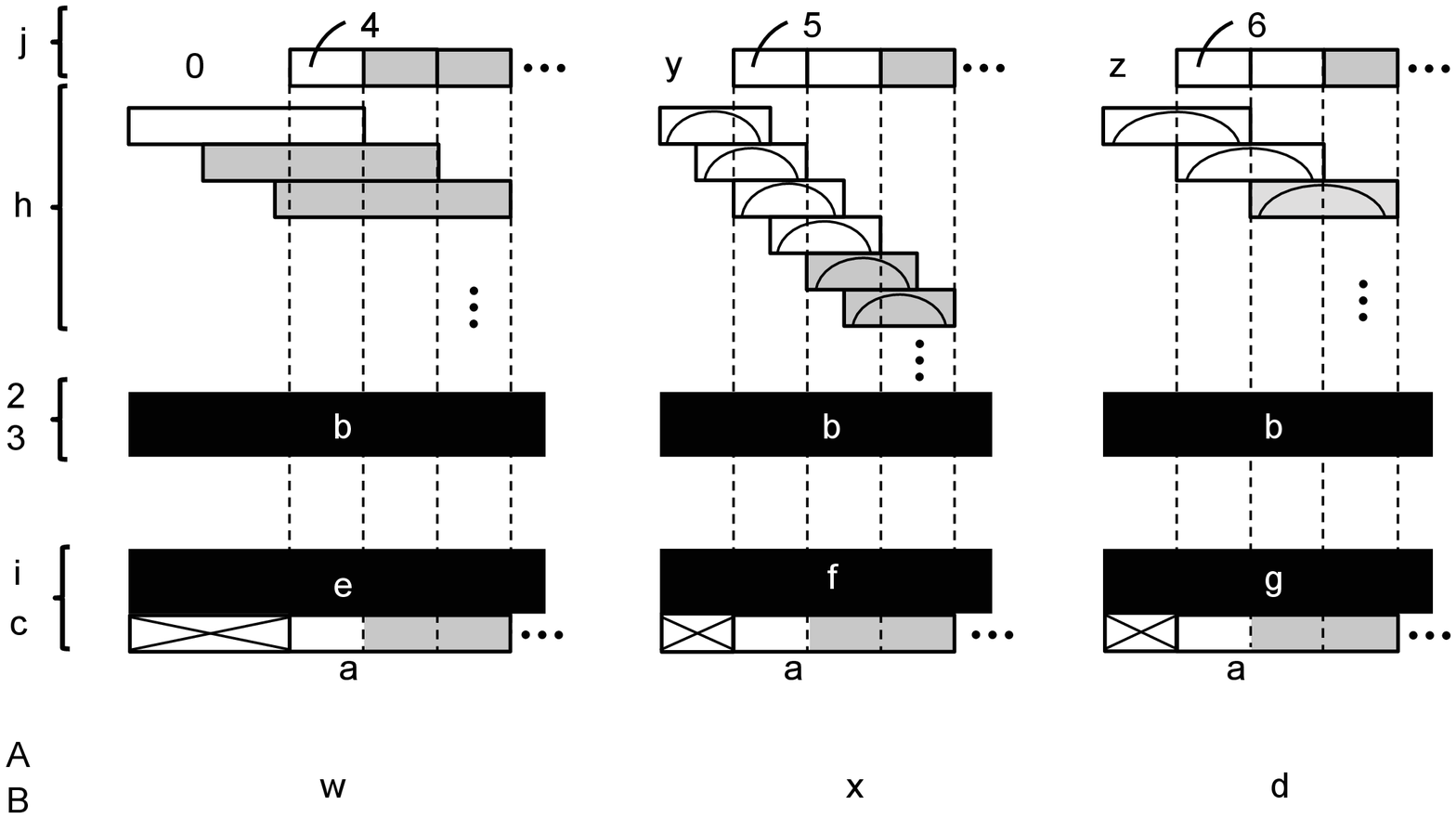}%
		\label{figures_postprocess_frame_structure1}}
	\hfil
	\subfloat{\includegraphics[width=0.5\textwidth]{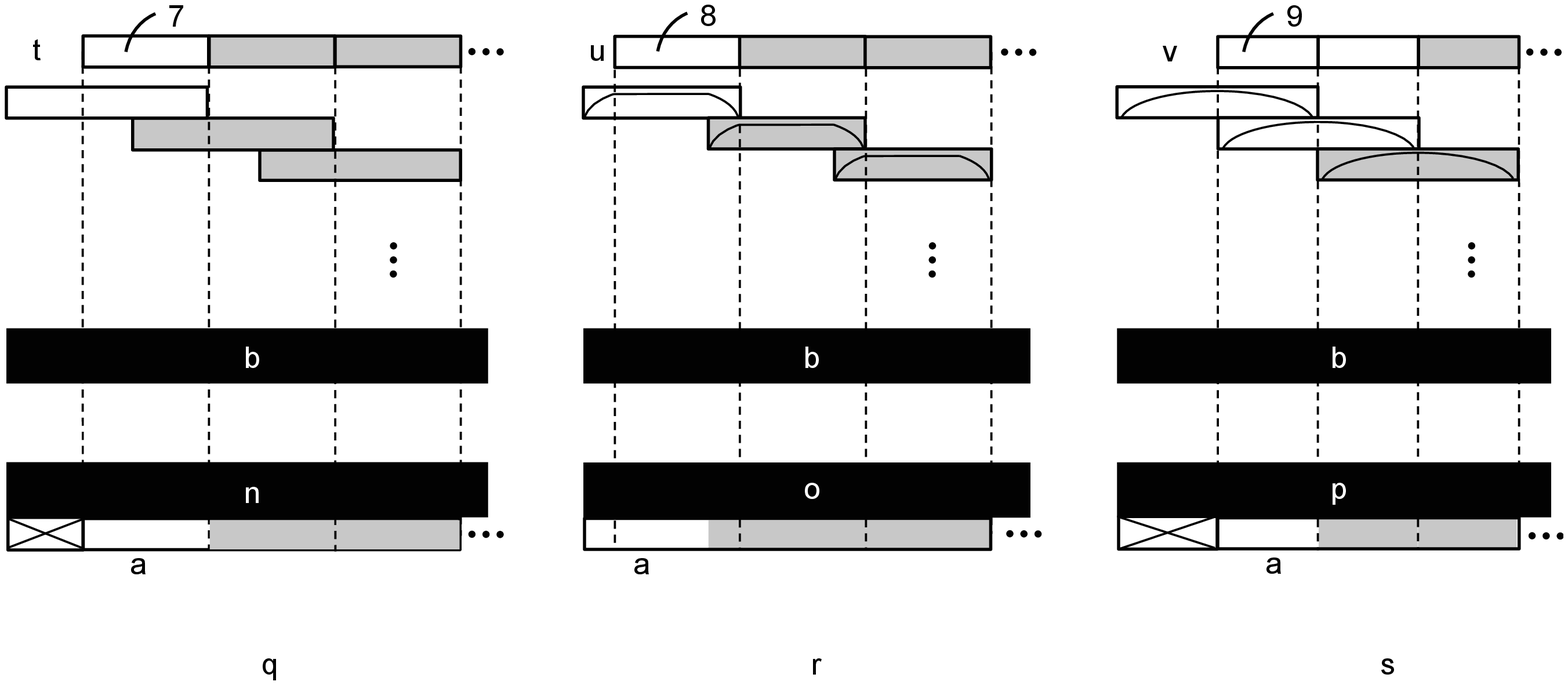}%
		\label{figures_postprocess_frame_structure2}}
	\caption{{\bf Framework structures} for windowing, ({\bf cepstral domain}) processing, and waveform reconstruction. In the upper part of the figure, all signal portions necessary to be available for computing the first frame $\ell$ of $\hat{s}(n)$ are marked as white boxes \framebox(16,5){}\,, as is the current output frame $\hat{s}(n),n\in\mathcal{N}_{\ell}$, in the bottom part of the figure. OLA stands for overlap-add of all upper-part white windowing boxes for current frame $\ell$.}
	\label{figures_postprocess_frame}
\end{figure*}

\vspace{-3pt}
\subsection{Time Domain Approach: Processing}
\label{subsec_time_dom}
For the time domain approach, we choose a quite straightforward framework structure (i.e., windowing and waveform reconstruction) which fits to most speech decoders: a 10ms rectangular window without overlapping. The windowed frame $\mathbf{s}(\ell)$ then serves directly as the input of the CNN with the target being $\tilde{\mathbf{s}}(\ell)$, which is the noise-free undistorted (uncoded) windowed speech frame. Details of the CNN topology will be presented in Section \ref{subsec_CNN_topo}.
After CNN processing, the enhanced frame $\hat{\mathbf{s}}_{\text{t}}(\ell)$ is directly concatenated to reconstruct the waveform $\hat{s}(n)$. The motivation of this end-to-end time domain approach is to learn a mapping from the coded speech frame to the undistorted speech frame via the CNN, exploiting the temporal redundancy in terms of speech signal correlation in the decoder, to directly enhance the waveform of the coded speech. Beyond framing, no additional algorithmic delay is incurred. Note that this allows effectively latency-free postfiltering if the frame size matches the frame size of the speech decoder or if it matches the voice-over-IP packet size.

\begin{table}[]
	\begin{center}
		\footnotesize
		\begin{tabular}{m{2.05cm}<{\centering} | m{0.5cm}<{\centering}  m{0.55cm}<{\centering}  m{0.6cm}<{\centering}  m{0.55cm}<{\centering}  m{0.55cm}<{\centering}  m{0.1cm}<{\centering}   m{0.0001cm}}
			\hline \\[-6pt]
			  & \multicolumn{6}{c}{Framework Structure} \\
			 \\[-6pt]
			 & \uppercase\expandafter{\romannumeral1} & \uppercase\expandafter{\romannumeral2} & \uppercase\expandafter{\romannumeral3} & \uppercase\expandafter{\romannumeral4} & \uppercase\expandafter{\romannumeral5} & \uppercase\expandafter{\romannumeral6} &\\
			 \hline
			Window length $N_{\text{w}}$ [ms] & 32 & 15 & 20 & 32 & 25 & 32 &\\[4pt]
			\hline
			Processing length [ms] & 32 & 16 & 32 & 32 & 32 & 32 &\\[4pt]
			\hline
			Processing shift $N_{\text{s}}$ [ms] & 10 & 5 & 10 & 20 & 20 & 16 &\\[4pt]
			\hline
			Output overlap ratio & 0 & 66.7\% & 50\% & 0 & 20\% & 50\% &\\[4pt]
			\hline
			Additional delay [ms] & 0 & 10 & 10 & 0 & 5 & 16 &\\[4pt]			
			\hline
		\end{tabular}
	\end{center}
	\caption{Detailed settings of the framework structures \protect\\ for the \textbf{ cepstral domain} approach. }
	\label{table_structures}
\end{table}

\vspace{-4pt}
\subsection{Cepstral Domain Approach: Framework Structures}
This subsection presents the various framework structures for the cepstral domain approach, shown in Fig. \ref{figures_postprocess_frame}. On the one hand, since FFT and discrete cosine transform (DCT) are performed in the cepstral domain approach to obtain the cepstral coefficients (explained in detail in Section \ref{ceps_dom_process}), an appropriate frame length and overlapping setting are important. On the other hand, since the postprocessor follows the speech decoder, the frame lengths of typical decoders are also taken into consideration to design the framework structures. As a result, we investigate six framework structures to offer broad selections for various possible application scenarios. These structures can be divided into three groups: Structures \uppercase\expandafter{\romannumeral1}, \uppercase\expandafter{\romannumeral2} and \uppercase\expandafter{\romannumeral3} are designed for codecs with 10 ms frames,  \uppercase\expandafter{\romannumeral4} and \uppercase\expandafter{\romannumeral5} are designed for codecs with 20 ms frames, while structure \uppercase\expandafter{\romannumeral6} is for delay-insensitive off-line usage with 16 ms frames, one frame lookahead, and 50\% overlap. 

First of all, windowing of the coded speech $s(n)$ is implemented to form frames for processing, which can be denoted as
\begin{equation}
\label{frame_forming}
\mathbf{s}(\ell)=\big\lbrack s\big((\ell\!-\!1)N_{\text{s}}\big), \ldots, s\big((\ell\!-\!1)N_{\text{s}}+N_{\text{w}}\!-\!1\big) \big\rbrack \circ  \mathbf{w},
\end{equation}
where $N_{\text{s}}$ is the frame shift, $N_{\text{w}}$ is the length of window function, $\mathbf{w}$ is the window function vector, and $\circ$ denotes the sample-wise multiplication. As shown in Fig. \ref{figures_postprocess_frame}, all six frameworks require a few initial zeros to be padded to the coded input speech. The detailed settings of the framework structures are listed in Table \ref{table_structures}. It is worth noting that if the processing length is longer that the window length, a zero-padding is performed also after windowing.

After processing of the windowed frames, the speech waveform needs to be reconstructed, which is also illustrated in Fig. \ref{figures_postprocess_frame}. In structure \uppercase\expandafter{\romannumeral1} and structure \uppercase\expandafter{\romannumeral4}, only the latest samples of the processed frame are kept and the other samples are dropped, which means that beyond framing (10 ms and 20 ms, respectively) no additional algorithmic delay coccus. If used in conjunction with speech decoders operating with this frame size, or if used in conjunction with, e.g., G.711, G.726, or G.722, assembled to 10 ms voice-over IP packets, the entire postprocessing is effectively free of algorithmic delay (as is the case in the time domain approach, cf.\ Section \ref{subsec_time_dom}). In structures \uppercase\expandafter{\romannumeral2}, \uppercase\expandafter{\romannumeral3} and \uppercase\expandafter{\romannumeral6}, since periodic Hann windows are employed, the processed frames overlap and need to be added after time alignment. As a result, additional algorithmic delay is introduced for each of these three structures. Structure \uppercase\expandafter{\romannumeral5} aims at low complexity by using a flat-top periodic Hann window with low overlap ratio. In this structure, the output signal will be delayed by only 5ms, i.e., the output starts with 5 ms of zeros.  

\begin{figure*}[!t]
	\psfrag{a}[cc][cc]{\footnotesize FFT}
	\psfrag{b}[cc][cc]{\footnotesize log$_{10}|\!\cdot\!|$}
	\psfrag{c}[cc][cc]{\footnotesize DCT-\uppercase\expandafter{\romannumeral2}}
	\psfrag{d}[cc][cc]{\footnotesize Separation}
	
	\psfrag{e}[cc][cc]{\footnotesize CNN}
	\psfrag{f}[cc][cc]{\footnotesize Combination}
	\psfrag{g}[cc][cc]{\footnotesize IDCT-\uppercase\expandafter{\romannumeral2}}
	\psfrag{h}[cc][cc]{\footnotesize $10^{(\cdot)}e^{j(\cdot)}$}
	\psfrag{i}[tc][tc]{\footnotesize $\text{IFFT}$}
	\psfrag{j}[cc][cc]{\footnotesize arg$(\cdot)$}
	
	\psfrag{k}[cl][cl]{\footnotesize Cepstral Domain Processing}
	
	\psfrag{l}[cc][cc]{\footnotesize $\mathbf{s}(\ell)$}
	\psfrag{m}[cc][cc]{\footnotesize $\mathbf{S}(\ell)$}
	\psfrag{n}[cc][cc]{\footnotesize $\mathbf{c}(\ell)$}
	\psfrag{o}[cc][cc]{\footnotesize $\mathbf{c}_{\text{env}}(\ell)$}
	\psfrag{t}[cc][cc]{\footnotesize $\mathbf{c}_{\text{res}}(\ell)$}
	\psfrag{u}[cc][cc]{\footnotesize $\boldsymbol{\alpha}(\ell)$}
	\psfrag{p}[cc][cc]{\footnotesize $\hat{\mathbf{c}}_{\text{env}}(\ell)$}
	\psfrag{q}[cc][cr]{\footnotesize $\hat{\mathbf{c}}(\ell)$}
	\psfrag{r}[cc][cc]{\footnotesize $\hat{\mathbf{S}}(\ell)$}
	\psfrag{s}[cc][rc]{\footnotesize $\hat{\mathbf{s}}_{\text{c}}(\ell)$}
	
	\centering
	\subfloat{\includegraphics[width=0.92\textwidth]{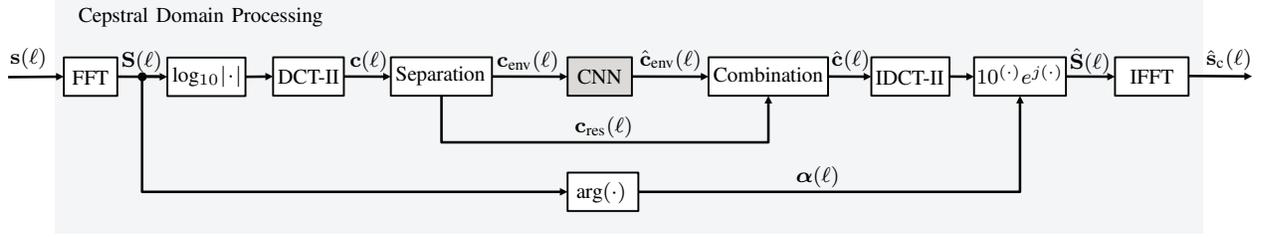}%
		}
	\vspace{-3pt}
	\caption{\textbf{Processing structure of the cepstral domain approach}. The topology of the CNN block is identical in the time domain and the cepstral domain approach and is depicted in Fig. \ref{figures_CNN_topology}.}
	\label{figures_postprocess_cepstral_dom}
\end{figure*}

\begin{figure}[!t]
	\psfrag{a}[cc][cc]{\footnotesize Conv$(F,\!N\!\times\! 1)$}
	\psfrag{b}[cc][cc]{\footnotesize Conv$(F,\!N\!\times\! F)$}
	\psfrag{c}[cc][cc]{\footnotesize Max Pooling$(2\!\times\! 1)$}
	\psfrag{d}[cc][cc]{\footnotesize Conv$(2F,\!N\!\times\! F)$}
	\psfrag{e}[cc][cc]{\footnotesize Conv$(2F,\!N\!\times\! 2F)$}
	
	\psfrag{f}[cc][cc]{\footnotesize Max Pooling$(2\!\times\! 1)$}
	\psfrag{g}[cc][cr]{\footnotesize Conv$(F,\!N\!\times\! 2F)$}
	\psfrag{h}[cc][cc]{\footnotesize Upsampling$(2\!\times\! 1)$}
	\psfrag{i}[cc][cc]{\footnotesize Conv$(2F,\!N\!\times\! F)$}
	\psfrag{j}[cc][cc]{\footnotesize Conv$(2F,\!N\!\times\! 2F)$}
	
	\psfrag{k}[cc][cr]{\footnotesize Upsampling$(2\!\times\! 1)$}
	\psfrag{l}[cc][cc]{\footnotesize Conv$(F,\!N\!\times\! 2F)$}
	\psfrag{m}[cc][cc]{\footnotesize Conv$(F,\!N\!\times\! F)$}
	\psfrag{n}[cc][cc]{\footnotesize Conv$(1,\!N\!\times\! F)$}
	
	\psfrag{o}[cc][cc]{\footnotesize $L\times 1$}
	\psfrag{p}[cc][cc]{\footnotesize $L\times F$}
	\psfrag{q}[cc][cc]{\footnotesize $L\times F$}
	\psfrag{r}[cc][cc]{\footnotesize $L/2\times F$}
	\psfrag{s}[cc][cc]{\footnotesize $L/2\times 2F$}
	\psfrag{t}[cc][cc]{\footnotesize $L/2\times 2F$}
	\psfrag{u}[cc][cc]{\footnotesize $L/4\times 2F$}
	\psfrag{v}[cc][cc]{\footnotesize $L/4\times F$}
	\psfrag{w}[cc][cc]{\footnotesize $L/2\times F$}
	\psfrag{x}[cc][cc]{\footnotesize $L/2\times 2F$}
	\psfrag{y}[cc][cc]{\footnotesize $L/2\times 2F$}
	\psfrag{z}[cc][cc]{\footnotesize $L\times 2F$}
	\psfrag{9}[cc][cc]{\footnotesize $L\times F$}
	\psfrag{8}[cc][rc]{\footnotesize $L\times F$}
	\psfrag{7}[cc][cc]{\footnotesize $L\times F$}
	\psfrag{6}[cc][cc]{\footnotesize $L\times 1$}
	
	\psfrag{A}[cc][cc]{\footnotesize $\mathbf{s}(\ell)$ or\ \  $\mathbf{c}_{\text{env}(\ell)}$}
	\psfrag{C}[cc][cc]{\footnotesize $\hat{\mathbf{s}}_{\text{t}}(\ell)$ or\ \  $\hat{\mathbf{c}}_{\text{env}(\ell)}$}
	
	\psfrag{5}[cc][cc]{\footnotesize CNN}
	
	\psfrag{4}[cc][cc]{\footnotesize (skip)}
	\psfrag{3}[cc][cc]{\footnotesize \rotatebox{90}{(skip)}}
	
	\centering
	\includegraphics[width=0.49\textwidth]{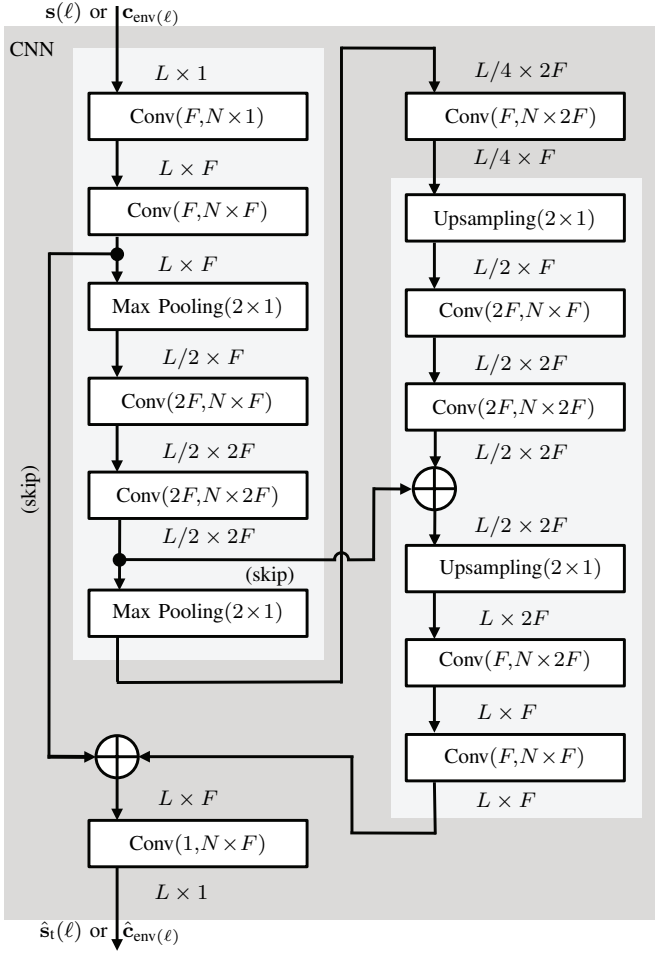}
	\vspace{-0.3cm}
	\caption{Detailed view of the {\bf CNN structure in both time domain} ($L$ equals 10 ms of speech samples) and {\bf cepstral domain} ($L\!=\!|\mathcal{M_{\text{env}}}|$). The operation Conv() stands for convolutional layers containing two parameters, which are the number of feature maps (filter kernels) $F$ or $2F$, and the kernel size ($a\!\times\!b$). The max pooling and upsampling layers are described by the kernel size ($2\!\times\! 1$). The input and output dimensions of each layer are also given. The light gray areas contain two symmetric procedures. }
	\label{figures_CNN_topology}
\end{figure}

\vspace{-3pt}
\subsection{Cepstral Domain Approach: Processing} \label{ceps_dom_process}
As we have learnt from the aforementioned formant postfilters, an emphasis of the spectral envelope peaks can reduce the impact of the coding distortion. By using cepstral domain envelope features, the dimension of the input vector to the CNN will be largely reduced compared to the time domain approach, which makes the CNN able to concentrate on the more perceptually relevant information, i.e., the formant structure. 

Our cepstral domain approach uses a CNN to restore the cepstral coefficients responsible for the spectral envelope and then synthesizes the speech frame using the enhanced envelope cepstral coefficients, as well as the residual cepstral coefficients and the phase information, the two latter both being acquired from the coded speech frame. The whole processing structure is shown in Fig. \ref{figures_postprocess_cepstral_dom}.

At first, the windowed frame is transformed to the frequency domain as vector $\mathbf{S}(\ell)$ using the $K$-point FFT. Subsequently, the cepstrum (cepstral coefficients) is computed by applying the discrete cosine transform of type \uppercase\expandafter{\romannumeral2} (DCT-\uppercase\expandafter{\romannumeral2}) to the logarithmic magnitude spectrum, which can be expressed as
\begin{equation}
\label{equ_ceps_coeff}
c(\ell,m)=\sum\limits_{k\in\mathcal{K}}\text{log}(\left | S(\ell,k) \right | )\cdot \text{cos}\big(\pi m(k+0.5)/K\big),
\end{equation}
where $k\!\in\!\mathcal{K}\!=\! \{0,\!\cdots\!,\!K\!-\!1\}$ is the frequency bin index and $m\!\in\!\mathcal{M} \!=\!\{0,\!1,\!\cdots\!,\!K\!-\!1\}$ is the index of cepstral coefficients. Then, the cepstrum is lowpass liftered (i.e., taking only the lower part of the cepstrum) to obtain the cepstral coefficients responsible for the spectral envelope, which is denoted as $c_{\text{env}}(\ell,m)$ with $m\!\in\!\mathcal{M}_{\text{env}}$. In this work, we regard the first $6.25\%$ cepstral coefficients as the coefficients responsible for the spectral envelope\textcolor{black}{\footnote{\textcolor{black}{As we have $K\!=\!512$ for narrowband speech, the $0.0625\!\cdot\! K\text{-th}\!=\!32$nd cepstral coefficient represents the frequency $1/(32\!\times\!\frac{1}{16}\text{ms})\!=\!500\text{Hz}$ (check (\ref{equ_ceps_coeff})\,!). Using 500Hz as cepstral lowpass liftering cutoff frequency, the fundamental frequency (F0) will be excluded in most cases. This is because the fundamental frequency can vary from 40 Hz for a very low-pitched male voice to 600 Hz for a very high-pitched female or child voice~\cite{huang2001spoken}. As a result, the pitch periodicity from speech is removed, while the information of the spectral envelope representing the formants is kept for further processing.}}}, resulting in $|\mathcal{M_{\text{env}}}|\!=\!6.25\%\!\cdot\!|\mathcal{M}|$. 

This vector $\mathbf{c}_{\text{env}}(\ell)$ serves as the input to the CNN, which then provides the restored cepstral coefficients responsible for the spectral envelope $\hat{\mathbf{c}}_{\text{env}}(\ell)$. 
After that, the residual cepstral coefficients from the liftering, denoted as $c_{\text{res}}(\ell,m)$ with $m\!\in\!\mathcal{M}_{\text{res}}$, are concatenated to $\hat{\mathbf{c}}_{\text{env}}(\ell)$ to constitute the complete cepstral coefficient vector $\hat{\mathbf{c}}(\ell)$. Then the logarithmic magnitude of the processed spectrum $\hat{\mathbf{S}}(\ell)$ is calculated by inverse DCT-\uppercase\expandafter{\romannumeral2} (IDCT-\uppercase\expandafter{\romannumeral2}) as
\begin{equation}
\begin{split}
\text{log}\!\left | \hat{S}(\ell,k) \right | \!  =\!  \frac{1}{K}\left [ \hat{c}(\ell,0)\!+\! 2\!\sum\limits_{m=1}^{K-1}\hat{c}(\ell,m)\!\cdot\!  \text{cos}\Big(\!\frac{\pi m(k+0.5)}{K}\!\Big)\right ]\!.
\end{split}
\end{equation}
Finally, the elements of $\hat{\mathbf{S}}(\ell)$ are subsequently obtained by
\begin{equation}
\hat{S}(\ell,k) = \left | \hat{S}(\ell,k) \right |  \text{exp}(j\cdot\alpha(\ell,k)),
\end{equation}
where $\boldsymbol{\alpha}(\ell)$ is the phase information from $\mathbf{S}(\ell)$. The processed frame $\hat{\mathbf{s}}_\text{c}(\ell)$ is obtained by performing the IFFT of $\hat{\mathbf{S}}(\ell)$. 

\begin{figure}[]
	\psfrag{a}[cc][tc]{\footnotesize Test}
	\psfrag{b}[cc][cc]{\footnotesize Set}
	\psfrag{c}[cc][tc]{\footnotesize Initial}
	\psfrag{d}[cc][Cc]{\footnotesize Filtering}
	\psfrag{e}[cc][tc]{\footnotesize Level}
	\psfrag{f}[cc][bc]{\footnotesize Adjustment}
	\psfrag{h}[cc][cc]{\footnotesize Concatenation}
	\psfrag{7}[bc][bc]{\footnotesize Down-}
	\psfrag{i}[cc][bc]{\footnotesize Sampling}
	\psfrag{j}[cc][tc]{\footnotesize Reference}
	\psfrag{k}[cc][cc]{\footnotesize Speech}
	\psfrag{l}[cc][cc]{\footnotesize Segmentation}
	\psfrag{m}[cc][tc]{\footnotesize Coded}
	\psfrag{n}[cc][cc]{\footnotesize Speech}
	
	\psfrag{o}[cc][cc]{\footnotesize Segmentation}
	\psfrag{p}[cc][cc]{\footnotesize DEC}
	\psfrag{q}[cc][bc]{\footnotesize Conversion}
	\psfrag{r}[cc][tc]{\footnotesize Bit\,Conver-}
	\psfrag{s}[cc][bc]{\footnotesize sion\&ENC}
	\psfrag{t}[cc][tc]{\footnotesize Enhanced}
	\psfrag{u}[cc][cc]{\footnotesize Speech}
	\psfrag{v}[cc][cc]{\footnotesize Segmentation}
	\psfrag{w}[cc][tc]{\footnotesize Post-}
	\psfrag{x}[cc][bc]{\footnotesize processor}
	\psfrag{y}[cc][tc]{\footnotesize DEC\&Bit}
	\psfrag{z}[cc][bc]{\footnotesize Conversion}
	\psfrag{9}[cc][tc]{\footnotesize Bit\,Conver-}
	\psfrag{8}[cc][bc]{\footnotesize sion\&ENC}
	
	\psfrag{H}[cc][cc]{\footnotesize VAD}
	
	\psfrag{J}[cc][cc]{\footnotesize Training\&}
	\psfrag{K}[cc][cc]{\footnotesize Validation}
	\psfrag{D}[tc][tc]{\footnotesize Target}
	\psfrag{L}[bc][bc]{\footnotesize Data}
	\psfrag{T}[tc][tc]{\footnotesize Preparation}
	\psfrag{M}[cc][cc]{\footnotesize Training\&}
	\psfrag{N}[cc][cc]{\footnotesize Validation}
	\psfrag{E}[tc][tc]{\footnotesize Input}
	\psfrag{O}[bc][bc]{\footnotesize Data}
	\psfrag{U}[tc][tc]{\footnotesize Preparation}
	
	\psfrag{P}[cc][cc]{\footnotesize DEC}
	\psfrag{Q}[cc][cc]{\footnotesize BitConv.\&}
	\psfrag{6}[tc][cc]{\footnotesize DelayComp.}
	
	\psfrag{R}[cc][cc]{\footnotesize DelayComp.}
	\psfrag{S}[cc][bc]{\footnotesize \&BitConv.}
	\psfrag{5}[cc][cc]{\footnotesize ENC}
	
	\psfrag{A}[cc][cc]{\footnotesize Training\&}
	\psfrag{B}[cc][cc]{\footnotesize Validation}
	\psfrag{C}[tc][tc]{\footnotesize Set}
	
	\psfrag{V}[cc][cc]{\footnotesize RMS Level}
	\psfrag{W}[bc][bc]{\footnotesize Noise}
	\psfrag{X}[tc][tc]{\footnotesize Data}
	\psfrag{Y}[cc][cc]{\footnotesize EID}
	\psfrag{Z}[cc][cc]{\footnotesize Noisy Condition}
	\psfrag{4}[cc][cr]{\footnotesize Error-Prone}
	\psfrag{3}[cc][cr]{\footnotesize Condition}
	
	\centering
	\vspace{-4pt}
	\includegraphics[width=0.5\textwidth]{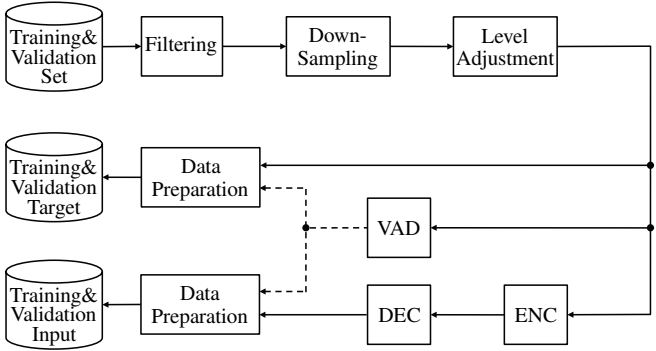}%
	\caption{{\bf Training} and {\bf validation} preprocessing.}
	\label{figures_process_plan1}
\end{figure}

\subsection{Both Approaches: CNN Topology}
The CNN topology, both in the time domain approach or in the cepstral domain approach, is a deep convolutional encoder-decoder network, which is shown in Fig. \ref{figures_CNN_topology}. This topology is motivated from~\cite{mao2016image} and three different kinds of layers are used in this topology which will be explained in the following. 

The \emph{convolutional layers} are defined by the number $F$ or $2F$ of feature maps (filter kernels) and the kernel size ($a\!\times\!b$). The number of trainable weights, including the bias, of a convolutional layer denoted as, e.g., the first layer (Conv$(F,\!N\!\times\! 1)$), results in $F\!\times\!(N\!\times\!1)\!+\!F$. It is worth noting that in each convolutional layer, the stride is 1 and zero-padding of the layer input is always performed to guarantee that the first dimension of the layer output is the same as that for the layer input. In \emph{max pooling layers}, a $2\!\times\!1$ maximum filter is applied in a non-overlapping fashion, resulting in a $50\%$ reduction of the layer input along the first dimension. On the contrary, the \emph{upsampling layers} simply copy each element of the layer input into a $2\!\times\!1$ vector and stack these vectors just following the original order, which actually doubles the first dimension of the layer input.

As can be seen in Fig. \ref{figures_CNN_topology}, two light gray areas include two symmetric procedures,  respectively. In the first procedure, the convolutional layers and the max pooling layers are used together to extract the relevant information and to discard the corrupted parts of the CNN input feature vector, resulting in a compression of the vector length. 
The second procedure is designed to recover the detail information via the combination of upsamping layers and convolutional layers. Meanwhile, the vector length is increased back to the original dimension by using two times the upsampling layer. In the last convolutional layer, a linear activation function is used and the final output has exactly the same dimension $L$ as the input of the CNN. Furthermore, two skip connections are utilized to add up the corresponding layer outputs, in order to ease the vanishing gradient problem during the training of this deep CNN~\cite{mao2016image}. 

\section{Experimental Setup and Metrics} \label{sec_exp_set_metric}
\subsection{Speech Database}
Speech data used in this work is from the NTT wideband speech database~\cite{NTT-AT_MultilingualTelephonometry}, 
containing 21 different languages, and 4 female and 4 male speakers for each language. Each of the speakers is represented by 12 speech utterances of about 8 seconds duration. American English and German are used for test, and for each language, the test set contains 30 speech utterances, in which 3 female speakers and the 3 male speakers are represented by 5 different speech utterances, respectively. For the training set, all speech utterances from 3 female speakers and 3 male speakers \emph{in all other 19 languages} are chosen, while 9 speech utterances from each of the remaining speakers (female speaker \textcolor{black}{\texttt{f4}} and male speaker \textcolor{black}{\texttt{m4}} per language) in the same 19 languages are used as validation set. Thereby we provide (partly\footnote{It should be mentioned that British English is one of the 19 training and validation languages, while American English is used in the test. The subjective listening test, however, will be conducted with German samples only, thus being completely language-independent.}) language-independent but completely speaker-independent results throughout.

\begin{figure}[]
	\psfrag{a}[cc][tc]{\footnotesize Test}
	\psfrag{b}[cc][cc]{\footnotesize Set}
	\psfrag{c}[cc][tc]{\footnotesize Initial}
	\psfrag{d}[cc][Cc]{\footnotesize Filtering}
	\psfrag{e}[cc][tc]{\footnotesize Level}
	\psfrag{f}[cc][bc]{\footnotesize Adjustment}
	\psfrag{h}[cc][cc]{\footnotesize Concatenation}
	\psfrag{7}[bc][bc]{\footnotesize Down-}
	\psfrag{i}[cc][bc]{\footnotesize Sampling}
	\psfrag{j}[cc][tc]{\footnotesize Reference}
	\psfrag{k}[cc][cc]{\footnotesize Speech}
	\psfrag{l}[cc][cc]{\footnotesize Segmentation}
	\psfrag{m}[cc][tc]{\footnotesize Coded}
	\psfrag{n}[cc][cc]{\footnotesize Speech}
	
	\psfrag{o}[cc][cc]{\footnotesize Segmentation}
	\psfrag{p}[cc][cc]{\footnotesize DEC}
	\psfrag{q}[cc][bc]{\footnotesize Conversion}
	\psfrag{r}[cc][tc]{\footnotesize Bit\,Conver-}
	\psfrag{s}[cc][bc]{\footnotesize sion\&ENC}
	\psfrag{t}[cc][tc]{\footnotesize Enhanced}
	\psfrag{u}[cc][cc]{\footnotesize Speech}
	\psfrag{v}[cc][cc]{\footnotesize Segmentation}
	\psfrag{w}[cc][tc]{\footnotesize Post-}
	\psfrag{x}[cc][bc]{\footnotesize processor}
	\psfrag{y}[cc][tc]{\footnotesize DEC\&Bit}
	\psfrag{z}[cc][bc]{\footnotesize Conversion}
	\psfrag{9}[cc][tc]{\footnotesize Bit\,Conver-}
	\psfrag{8}[cc][bc]{\footnotesize sion\&ENC}
	
	\psfrag{H}[cc][cc]{\footnotesize VAD}
	
	\psfrag{J}[cc][cc]{\footnotesize Training\&}
	\psfrag{K}[cc][cc]{\footnotesize Validation}
	\psfrag{D}[tc][tc]{\footnotesize Target}
	\psfrag{L}[bc][bc]{\footnotesize Data}
	\psfrag{T}[tc][tc]{\footnotesize Preparation}
	\psfrag{M}[cc][cc]{\footnotesize Training\&}
	\psfrag{N}[cc][cc]{\footnotesize Validation}
	\psfrag{E}[tc][tc]{\footnotesize Input}
	\psfrag{O}[bc][bc]{\footnotesize Data}
	\psfrag{U}[tc][tc]{\footnotesize Preparation}
	
	\psfrag{P}[cc][cc]{\footnotesize DEC}
	\psfrag{Q}[cc][cc]{\footnotesize BitConv.\&}
	\psfrag{6}[tc][cc]{\footnotesize DelayComp.}
	
	\psfrag{R}[cc][cc]{\footnotesize DelayComp.}
	\psfrag{S}[cc][bc]{\footnotesize \&BitConv.}
	\psfrag{5}[cc][cc]{\footnotesize ENC}
	
	\psfrag{A}[cc][cc]{\footnotesize Training\&}
	\psfrag{B}[cc][cc]{\footnotesize Validation}
	\psfrag{C}[tc][tc]{\footnotesize Set}
	
	\psfrag{V}[bc][bc]{\footnotesize RMS Level}
	\psfrag{W}[bc][bc]{\footnotesize Noise}
	\psfrag{X}[tc][tc]{\footnotesize Data}
	\psfrag{Y}[cc][cc]{\footnotesize EID}
	\psfrag{Z}[cc][cc]{\footnotesize Noisy Condition}
	\psfrag{4}[cc][tr]{\footnotesize Error-Prone}
	\psfrag{3}[cc][cr]{\footnotesize Condition}
	
	\centering
	\vspace{-5pt}
	\includegraphics[width=0.5\textwidth]{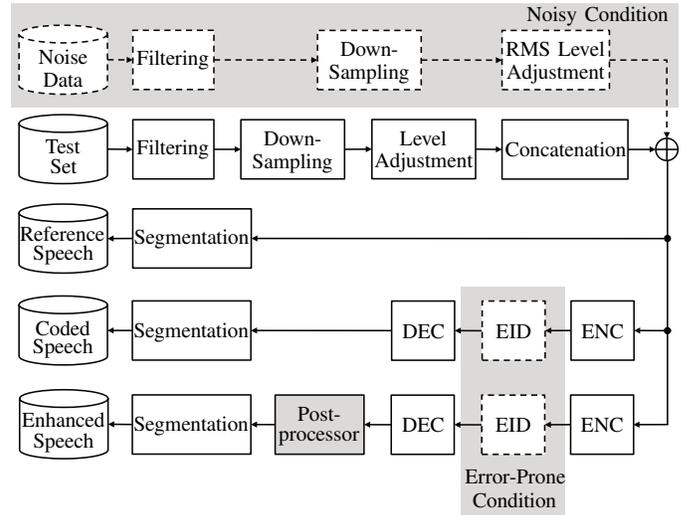}%
	\caption{{\bf Test} processing for various codecs and postprocessors in clean, error-prone transmission, and noisy conditions.} 
	\label{figures_process_plan2}
\end{figure}

\vspace{-3pt}
\subsection{Preprocessing for Training and Validation}
The training and validation data pairs (i.e., input and target) are obtained following the training and validation preprocessing illustrated in Fig.~\ref{figures_process_plan1}, and the test experiments follow the test processing in Fig.~\ref{figures_process_plan2}. Our training and validation preprocessing and test processing are based upon the original quality assessment plans~\cite{Plan_ITUTG711,Plan_AMRWB_NB,ramo2005comparing,Plan_EVS} for the codecs evaluated in this work and the respective processing functions employed in Figs.~\ref{figures_process_plan1} and \ref{figures_process_plan2} are from the ITU-T software tool library G.191~\cite{ITU_G191}. 

The speech utterances are firstly processed by different filters (i.e., FLAT for narrowband codecs\textcolor{black}{\footnote{\textcolor{black}{Note that for bandwidth consistency reasons, we decided to use the FLAT filter also for G.726 transmission, although typically here an MSIN filter response is used~\cite{ramo2005comparing,Plan_EVS}.}}} and P.341 for wideband codecs). Then, for narrowband codecs the speech signal is decimated from 16 kHz to 8 kHz using the high quality finite impulse response (FIR) low-pass filter HQ2 from~\cite{ITU_G191}, while for wideband codecs this downsampling function is bypassed. Then, the active speech level is adjusted to -26 dBov~\cite{ITUT_P56}. 

After this, to obtain the frame indices for the training and validation a very simple frame-based voice activity detection (VAD) is executed as 
\begin{equation}
\label{equ_vad}
\text{VAD}(\ell)=
\begin{cases}
1, ~\text{if} ~\frac{\frac{1}{|\mathcal{N_{\ell}}|}\sum_{n\in\mathcal{N_{\ell}}} \tilde{s}^2(n)}{\frac{1}{|\mathcal{N}|}\sum_{n\in\mathcal{N}} \tilde{s}^2(n)}>\theta_{\text{VAD}} \\
0, ~\text{else},
\end{cases}
\end{equation}
where $\theta_{\text{VAD}}$ is the VAD threshold, $\mathcal{N_{\ell}}$ and $\mathcal{N}$ are the sets of sample indices belonging to frame $\ell$ and the whole speech file, respectively. The frames marked with $\text{VAD}(\ell)\!=\!1$ are regarded as active speech frames and the corresponding frame indices are denoted as a set $\mathcal{L}_\text{VAD}\!=\!\left \{ \ell\!\mid\! \text{VAD}(\ell)\!=\!1 \right \}$. These active speech frames are further used for training and validation, while the other frames are regarded as speech pause and not used in this stage. 
Then, the target and input for training and validation are obtained as follows:

The \emph{target data} is obtained after the data preparation, in which the windowing w.r.t.\ the selected time domain or cepstral domain approach is applied to the active speech frames $\ell\!\in\!\mathcal{L}_\text{VAD}$.

For the \emph{input data of training and validation}, the level-adjusted speech is subject to coding. We examine in total four different speech codecs: two narrowband codecs, which are G.711~\cite{ITUTG711} and the adaptive differential pulse-code modulation (ADPCM) Recommendation G.726 used for digital enhanced cordless telephony (DECT) at 32 kbps~\cite{ITU_G726}, two wideband codecs, which are the wideband ADPCM G.722 used for wideband DECT at 64 kbps~\cite{ITU_G722}, and AMR-WB at 12.65 kbps~\cite{AMRWB3GPP} in \textcolor{black}{fixed-point implementation~\cite{AMRWB3GPP_fixedpointC} without DTX}. The function ``ENC'' comprises a delay compensation function in case of wideband codecs (cf.\ assessment plan~\cite{Plan_EVS}), a bit conversion function from 16 bits to 14 bits (only for wideband codecs) and the speech encoder from any of the above four codecs. Then, the corresponding function ``DEC'' is conducted, which comprises the speech decoder, a bit conversion function from 16 bits to 14 bits (only for wideband codecs), and a delay compensation function (only for wideband codecs). Finally, the coded frames with $\ell\!\in\!\mathcal{L}_\text{VAD}$ form the input data to the data preparation function, which again performs windowing and potential transformation to the cepstral domain. 

\vspace{-3pt}
\subsection{Processing for Training and Validation}
In the training processing, \textcolor{black}{we {\it always} train codec-individual CNN models which are then used later on in test.} The prepared input data in the respective domain according to Fig. \ref{figures_process_plan1} is at first normalized towards zero mean and unit variance, then this normalized input data and target data is fed into the CNN to train the weights in each convolutional layer. This is achieved by minimizing the cost function, which is the mean squared error (MSE) between the outputs of the CNN and the target data. Instead of using the traditional stochastic gradient descent (SGD) algorithm for the trainable weights updating, Adam~\cite{Kingma2015adam} is used as the learning method to obtain a faster training convergence~\cite{mao2016image}. In this work, the weights update is performed in each minibatch consisting of 16 frames, being a good trade-off between training speed and performance. At the beginning of each epoch, the training data is shuffled so that the 16 frames of each minibatch are randomly selected from the training data. 

In order to train the CNN in an efficient way and to avoid overfitting, the strategies for the learning rate and the stop criteria are the following: The initial learning rate is $5\!\times\!10^{-4}$ and it is halved once the MSE on the validation set does not decrease for two epochs. The training stop criterion is checked after each epoch, i.e., after all minibatches have been used, and the training stops if either the MSE on the validation set does not decrease for 16 epochs, or if the number of epochs approaches 100. Finally, the weights are saved as the result of that epoch, after which the lowest MSE on the validation set has been achieved. 

\vspace{-3pt}
\subsection{Processing for Test}
In Fig.~\ref{figures_process_plan2}, the test processing functions of filtering, downsampling, level adjustment, ``ENC'' and ``DEC'' are identical to those in Fig.~\ref{figures_process_plan1}. Since the proposed postprocessing approaches are evaluated in four conditions, i.e., clean, noisy, tandeming, and error-prone transmission conditions, the test processing is also described for these four conditions. \textcolor{black}{We always select the CNN model that refers to the last employed speech decoder. In most practical applications the last decoder can be assumed to be known, even if in many cases tandem conditions are observed with G.711 being such ``last employed decoder''. Please note that for the sake of conciseness, we did not include in our simulation the condition when the last decoder is unknown; this could be practically solved by a multi-codec-trained CNN model.}

For the \emph{clean condition}, the level-adjusted speech utterances are concatenated to a long speech signal, in which the utterances from female and male speakers are alternately concatenated. After this, the reference speech, coded speech and enhanced speech are obtained as follows: 

The reference speech is obtained after segmentation, which cuts the concatenated speech signal back to the original signal portions/durations. Note that this reference speech is also used for the other three conditions.  

To obtain coded speech, the function ``ENC'' and ``DEC'' are conducted and then the coded speech results after segmentation. 

To obtain enhanced speech, the functions ``ENC'' and ``DEC'' are conducted and then any of the postprocessors afterwards. Finally the enhanced speech files results after segmentation.

In the \emph{noisy conditions}, three types of noise from the ETSI background noise database~\cite{ETSI_noise_data} are applied in the evaluation part, which are cafeteria noise, car noise at the velocity of 100 km/h, and outside traffic road noise. Similar to the processing of speech utterances in Fig. \ref{figures_process_plan2}, the noise data is filtered and downsampled or bypassed depending on the codec bandwidth. Then the root mean square (RMS) level of noise is adjusted based on the desired SNR in dB~\cite{ITUT_P56}. After this, the adjusted noise is added to the concatenated speech for further processing. Finally, the coded and enhanced speech in the noisy condition are obtained with the same functions as in the clean condition. 

In \emph{error-prone transmission conditions}, e.g., mobile and wireless systems, frame losses are inserted to the bitstream after the encoder by using error insertion device (EID)~\cite{ITU_G191}, which is placed between the ``ENC'' and ``DEC'' in Fig.~\ref{figures_process_plan2}. The coded and enhanced speech in the error-prone transmission conditions are obtained with the other functions being the same as in the clean condition. Two kinds of frame losses are taken into consideration: random frame erasure, which is based on a Gilbert model and burst frame erasure, in which the occurrence of the bursts is modeled by the Bellcore model~\cite{ITU_G191,gilbert1960capacity}. Both kinds of frame erasures are characterized by the frame erasure ratio (FER), which is the ratio of the number of distorted frames vs. the number of all transmitted frames.

In \emph{tandeming conditions} we employ a receiver-sided postprocessor for G.711 A-law (narrowband) or the AMR-WB, with various previously mentioned codecs as former codecs, but also the narrowband AMR codec at 12.2 kbps~\cite{AMR3GPP}, wideband codecs G.711.1 with mode R3 at 96 kbps~\cite{ITUTG7111NEW}, and EVS-WB at 13.2 kbps~\cite{EVS3GPP}. The ``EID'' block in Fig.\ \ref{figures_process_plan2} is simply replaced by ``DEC'' and the subsequent ``ENC'', resulting in a serial connection of two codecs. 

\begin{table}[]
	\begin{center}
		\footnotesize
		\begin{tabular}{m{0.02cm}<{\raggedright}  m{0.09cm}<{\centering} | m{0.52cm}<{\centering}  m{0.55cm}<{\centering} | m{0.52cm}<{\centering}  m{0.55cm}<{\centering} | m{0.52cm}<{\centering}  m{0.55cm}<{\centering} | m{0.52cm}<{\centering}  m{0.55cm}<{\centering} m{0.001cm}}
			\cline{1-10} \\[-6pt]
			& \multicolumn{9}{c}{Number of feature maps $F$} & \\
			\multicolumn{2}{c}{}   & \multicolumn{2}{c}{20} & \multicolumn{2}{c}{22} & \multicolumn{2}{c}{24} & \multicolumn{2}{c}{26} & \\ \cline{1-10} 
			&   & Leaky ReLU & SELU & Leaky ReLU & SELU & Leaky ReLU & SELU &Leaky ReLU &  SELU &\\[12pt]
			\cline{3-10}
			& 2    & 10.77 & 10.98 & 10.79 & 10.76 & 10.57 & 10.44 & 10.74 & 10.53  & \\[3pt]
			\multirow{2}{*}{$N$} 
			& 4    & 8.50 & 8.65 & 8.54 & 8.44 & 8.37 & 8.65 & 8.53 & 8.52 & \\[3pt]
			
			& 6    & 8.30 & 8.44 & \bf 8.29 & 8.61 & 8.30 & 8.42 & 8.39 & 8.30 & \\[3pt]
			
			& 8    & 8.38 & 8.44 & 8.46 & 8.56 & 8.45 & 8.50 & 8.41 & 8.47 & \\[3pt]
			
			& 10   & 8.46 & 8.37 & 8.49 & 8.44 & 8.41 & 8.38 & 8.41 & 8.50 &  \\[3pt] 
			\cline{1-10}
		\end{tabular}
	\end{center}
	\caption{\textbf{Mean logarithmic spectral distance ($\text{LSD}$) [dB]} on the validation set. The best setting is written in \textbf{bold face}.}
	\label{tab_prelim}
\end{table}

\vspace{-5pt}
\subsection{Metrics of Speech Quality}
To instrumentally evaluate the enhanced speech $\hat{s}(n)$, the mean logarithmic spectral distance (LSD) averaged over frames is employed~\cite{abel2017instrumental}. The LSD is calculated as 
\begin{equation}
\label{equ_lsd}
\text{LSD}(\ell)=\sqrt{\!\frac{1}{k_{\text{high}}-k_{\text{low}}}\!\sum\limits_{k=k_{\text{low}}}^{k_{\text{high}}}\!\left [\! 10\text{log}_{10}\!\left (\! \tfrac{\left | \tilde{S}(\ell,k) \right |^2}{\left | \hat{S}(\ell,k) \right |^2} \!\right ) \!\right ]^2},
\end{equation}
where $\tilde{S}(\ell,k)$ and $\hat{S}(\ell,k)$ and the $k$-th FFT coefficient of the uncoded and the processed (can be either coded or postprocessed) speech signal in frame $\ell$, respectively, and $k_{\text{high}}$ and $k_{\text{low}}$ are the indices of the upper and lower frequency bin bounds taken into account. 
The frames used for the mean LSD are from the active speech frame set $\mathcal{L}_\text{VAD}$ (from equation (\ref{equ_vad})) and each frame is formed by employing a 32 ms periodic Hann window with 50\% overlap.

\begin{table}[]
	\begin{center}
		\footnotesize
		\setlength\tabcolsep{3.5pt} 
		{\color{black}\begin{tabular}{m{1.5cm}<{\centering} | m{0.5cm}<{\centering} | m{0.8cm}<{\centering} | m{0.6cm}<{\centering} | m{1.47cm}<{\centering} | m{0.5cm}<{\centering} | m{0.8cm}<{\centering}|  m{0.6cm}<{\centering} m{0.0001cm}} 
			\cline{1-8}
			\multicolumn{4}{c|}{Time Domain} & \multicolumn{4}{c}{Cepstral Domain} &\\[4pt]
			Topology & $r$ & \# of Param. & LSD [dB] & Topology & $r$ & \# of Param. & LSD [dB] &\\[4pt]
			\cline{1-8}
			CNN & \multirow{2}{*}{-} & \multirow{2}{*}{\bf 0.82M} & \multirow{2}{*}{\bf 11.03} & 	CNN  & \multirow{2}{*}{-} & \multirow{2}{*}{\bf 52.82K} & \multirow{2}{*}{\bf 8.29} &\\[3pt]
			$(F_{\text{opt}},N_{\text{opt}})$ & & & & $(F_{\text{opt}},N_{\text{opt}})$ & & & &\\
			\cline{1-8}
			\multirow{2}{*}{1024-1024}  & 0 & \multirow{2}{*}{1.21M} & 12.12 & 	\multirow{2}{*}{256-256}  & 0 & \multirow{2}{*}{82.46K} & 9.59 &\\[3pt]
			& 0.1 &   & 14.71 &    & 0.1 &   & 9.54 &\\
			\cline{1-8}
			\multirow{2}{*}{512-512-1024}  & 0 & \multirow{2}{*}{0.91M} & 12.33 & 	\multirow{2}{*}{128-128-256}  & 0 & \multirow{2}{*}{61.98K} & 9.89 &\\[3pt]
			& 0.1 &   & 28.90 &    & 0.1 &   & 9.63 &\\
			\cline{1-8}
			512-512-512  & 0 & \multirow{2}{*}{0.87M} & 13.08 & 128-128-128  & 0 & \multirow{2}{*}{57.89K} & 9.70 &\\[3pt]
			-512 & 0.1 &   & 30.68 &  -128  & 0.1 &   & 11.14 &\\
			\cline{1-8}
			512-512-256  & 0 & \multirow{2}{*}{0.87M} & 12.03 & 128-128-64  & 0 & \multirow{2}{*}{57.95K} & 9.24 &\\[3pt]
			-512-512 & 0.1 &   & 55.03 &  -128-128  & 0.1 &   & 15.53 &\\
			\cline{1-8}
		\end{tabular}}
	\end{center}
	\textcolor{black}{
		\caption{\textbf{$\text{LSD}$ [dB]} values on the validation set and the number of trainable parameters (\# of Param.) for the optimal CNN with $(F_{\mathrm{opt}},N_{\mathrm{opt}})$ and four different fully-connected neural networks with or without dropout (dropout rate $r$) in time domain and cepstral domain. The topologies yielding the lowest LSD values are written in \textbf{bold face}.}
		}
	\label{tab_prelim_fcnn}
\end{table}

To measure the speech distortion, the segmental speech-to-speech-distortion ($\text{SSDR}_{\text{seg}}$)~\cite{elshamy2017instantaneous} is calculated as 
\begin{equation}
\label{equ_ssdr}
\text{SSDR}_{\text{seg}}=\frac{1}{\left|\mathcal{L}_\text{VAD}\right|}\sum_{\ell\in\mathcal{L}_\text{VAD}}\text{SSDR}(\ell),
\end{equation} 
where $\text{SSDR}(\ell)$ is limited from $R_{\text{min}}\!=\!-10$ dB to $R_{\text{max}}\!=\!40$ dB by $\text{SSDR}(\ell)\!=\!\text{max}\left \{ \text{min}\left \{ \text{SSDR}'(\ell),\!R_{\text{max}} \right \}\!,\! R_{\text{min}} \right \}$. The term $\text{SSDR}'(\ell)$ is actually calculated as 
\begin{equation}
\label{equ_ssdr_prim}
\text{SSDR}'(\ell)=10\text{log}_{10}\left [ \tfrac{\sum_{n\in\mathcal{N}_{\ell}}\tilde{s}^2(n)}{\sum_{n\in\mathcal{N}_{\ell}}\left ( \hat{s}(n)-\tilde{s}(n) \right )^2} \right ],
\end{equation}
where $\mathcal{N}_{\ell}$ is the set of sample indices $n$ belonging to frame $\ell$, $\tilde{s}(n)$ and $\hat{s}(n)$ are the uncoded and time-aligned processed (can be either coded or postprocessed) speech signal, respectively. Each frame is also 32 ms with 50\% overlap. 
\textcolor{black}{
	Note that at some point we will also report on a {\it global} SSDR measure, which is simply obtained by (\ref{equ_ssdr_prim}) with setting $\mathcal{N}_{\ell}\!=\!\mathcal{N}$, meaning that all samples in each file contribute to each of the sums in (\ref{equ_ssdr_prim}). We will call this measure simply SSDR.
}

For instrumental assessment of speech quality, perceptual evaluation of speech quality (PESQ)~\cite{ITUT_pesq,ITUT_pesq_mapping} for the narrowband speech and WB-PESQ~\cite{ITUT_pesq_wb} for the wideband speech are used. The output of the two metrics is the mean opinion score (MOS) listening quality objective (LQO), which is denoted as MOS-LQO. A mean value over all test speech utterances for each respective language is reported in the evaluation. 
\textcolor{black}{In addition to (WB-)PESQ, we also perform perceptual objective listening quality prediction (POLQA)~\cite{ITUT_polqa_2018}. This is done in only a few conditions, checking, whether both measures lead to similar conclusions. 
	}

\begin{table*}[!htb] 
	\begin{center}
		\footnotesize
		\begin{tabular}{m{0.001cm}<{\centering} m{1.8cm}<{\centering} m{1.6cm}<{\centering} | m{1.2cm}<{\centering}  m{0.6cm}<{\centering}  m{0.7cm}<{\centering}  m{0.7cm}<{\centering} m{1.1cm}<{\centering} | m{1.2cm}<{\centering}  m{0.6cm}<{\centering}  m{0.7cm}<{\centering} m{0.7cm}<{\centering} m{1.1cm}<{\centering} m{0.001cm}}
			\cline{1-13}
			& & & \multicolumn{5}{c|}{American English} & \multicolumn{5}{c}{German}& \\[3pt]
			& & & \multicolumn{2}{c}{G.711 A-law} & \multirow{2}{*}{G.726} & \multirow{2}{*}{G.722} & \multirow{2}{*}{AMR-WB} & \multicolumn{2}{c}{G.711 A-law} & \multirow{2}{*}{G.726} & \multirow{2}{*}{G.722} & \multirow{2}{*}{AMR-WB}&\\
			
			& & & no Constr. & Constr. & & & & no Constr. & Constr. & & & & \\
			\cline{1-13}
			& \multirow{2}{*}{Legacy Codec} & \multirow{2}{*}{MOS-LQO} & \multicolumn{2}{c}{\multirow{2}{*}{\,\,\,\,\,\,\,4.21}} & \multirow{2}{*}{3.96} & \multirow{2}{*}{3.72} & \multirow{2}{*}{3.60} & \multicolumn{2}{c}{\multirow{2}{*}{\,\,\,\,\,\,\,4.15}} & \multirow{2}{*}{4.01} & \multirow{2}{*}{3.61} & \multirow{2}{*}{3.53} &\\
			& & & & & & & & & & & & &\\
			\cline{1-13}
			& \multirow{2}{*}{Postfilter~\cite{ITUTG711tollbox}} & MOS-LQO &    \multicolumn{2}{c}{\,\,\,\,\,\,\,4.32} & \multirow{2}{*}{-} & \multirow{2}{*}{-} & \multirow{2}{*}{-} & \multicolumn{2}{c}{\,\,\,\,\,\,\,4.25} & \multirow{2}{*}{-} & \multirow{2}{*}{-} & \multirow{2}{*}{-} &\\[3pt]
			& & $\Delta$MOS-LQO & \multicolumn{2}{c}{\,\,\,\,\,\,\,0.11} &   &   &   & \multicolumn{2}{c}{\,\,\,\,\,\,\,0.10} &   &   &   &\\[3pt]
			\cline{1-13}
			& \multirow{2}{*}{Time Domain} & MOS-LQO & 4.32 & 4.32  & 4.21 & 4.32 & 3.61 & 4.30 & 4.30 & 4.26 & 4.29 & 3.62 &\\[3pt]
			& & $\Delta$MOS-LQO & 0.11 & 0.11 & 0.25 & 0.60 & 0.01 & 0.15 & 0.15 & 0.25 & 0.68 & 0.09 &\\[3pt]
			\cline{1-13}
			& \multirow{2}{*}{\uppercase\expandafter{\romannumeral1}} & MOS-LQO & 4.24 & 4.27 & 3.99 & 4.13  & 3.45  & 4.13 & 4.18 & 4.01 & 4.07 & 3.29 &\\[3pt]
			& & $\Delta$MOS-LQO & 0.03 & 0.06 & 0.03 & 0.41 & -0.15  & -0.02 & 0.03 & 0 & 0.46 & -0.24 &\\[3pt]
			\cline{3-13}
			& \multirow{2}{*}{\uppercase\expandafter{\romannumeral2}} & MOS-LQO & 4.40 & 4.30 & 4.15 & \bf 4.47 & 3.78  & 4.39 & 4.24 & 4.26 & 4.46 & 3.73 &\\[3pt]
			& & $\Delta$MOS-LQO & 0.19 & 0.09 & 0.19 & \bf 0.75 & 0.18 & 0.24 & 0.09 & 0.25 & 0.85 & 0.20 &\\[3pt]
			\cline{3-13}
			\multirow{6}{*}{\rotatebox{90}{Cepstral Domain}} & \multirow{2}{*}{\uppercase\expandafter{\romannumeral3}} & MOS-LQO & \bf 4.43 & 4.33 & 4.20 & \bf 4.47 & \bf 3.79 & \bf 4.42 & 4.26  & 4.29 & \bf 4.48 & \bf 3.74 &\\[3pt]
			& & $\Delta$MOS-LQO & \bf 0.22 & 0.12  & 0.24 & \bf 0.75 & \bf 0.19 & \bf 0.27 & 0.11 & 0.28 & \bf 0.87 & \bf 0.21 &\\[3pt]
			\cline{3-13}
			& \multirow{2}{*}{\uppercase\expandafter{\romannumeral4}} & MOS-LQO & 4.27 & 4.27 & 4.01 & 4.17 & 3.52 & 4.17 & 4.19 & 4.04 & 4.12  & 3.41 &\\[3pt]
			& & $\Delta$MOS-LQO & 0.06 & 0.06 & 0.05 & 0.45 & -0.08 & 0.02 & 0.04 & 0.03 & 0.51 & -0.12 &\\[3pt]
			\cline{3-13}
			& \multirow{2}{*}{\uppercase\expandafter{\romannumeral5}} & MOS-LQO & 4.42 & 4.31  & \bf 4.21 & 4.45 & 3.74 & 4.41  & 4.26 & \bf 4.30 & 4.44 & 3.67 &\\[3pt]
			& & $\Delta$MOS-LQO & 0.21 & 0.10 & \bf 0.25 & 0.73 & 0.14 & 0.26 & 0.11 & \bf 0.29 & 0.83 & 0.14 &\\[3pt]
			\cline{3-13}
			& \multirow{2}{*}{\uppercase\expandafter{\romannumeral6}} & MOS-LQO & \bf 4.44 & 4.31 & \bf 4.25 & \bf 4.50 & \bf 3.85 &  \bf 4.42 & 4.23 & \bf 4.32 & \bf 4.47 & \bf 3.79 &\\[3pt]
			& & $\Delta$MOS-LQO & \bf 0.23 & 0.10 & \bf 0.29 \bf& \bf 0.78 & \bf 0.25 & \bf 0.27 & 0.08 & \bf 0.31 & \bf 0.86 & \bf 0.26 &\\[3pt]
			\cline{1-13}
		\end{tabular}
	\end{center}
	\caption{MOS-LQO (\textbf{PESQ} and \textbf{WB-PESQ}) for legacy codecs and codecs with various postprocessors. The \textbf{top two} results in each column are written in \textbf{bold face}.}
	\label{tab_dif_codecs}
\end{table*}

In addition, for the most promising approaches, we conduct an semi-formal comparison category rating (CCR) subjective listening test according to the ITU-T Recommendation P.800~\cite{ITU_P800}. In a CCR test, a pair of two speech samples is presented to the listeners, and the quality judgment of the second sample compared to that of the first is made and rated on the comparison MOS (CMOS) scale ranging from -3 (much worse) to +3 (much better). 

\section{Experimental Evaluation and Discussion} \label{sec_exp_eva_dis}
In Section \ref{subsec_CNN_topo}, a preliminary experiment is implemented to investigate the CNN topology on the validation set. Then, the optimal setting will be used for the subsequent experiments. 

\vspace{-3pt}
\subsection{Preliminary Experiment on CNN Parameters} \label{subsec_CNN_topo}
In a preliminary experiment the optimal CNN topology settings with the framework structure \uppercase\expandafter{\romannumeral3} of the cepstral domain approach for G.711 postprocessing are selected. The number of feature maps $F$, the length of the CNN kernels $N$, and the activation function (the last layer is always linear) are examined. We investigate both leaky rectified linear unit (ReLU)~\cite{maas2013rectifier} and scaled exponential linear unit (SELU)~\cite{klambauer2017self}. Since narrowband speech is used in this preliminary experiment, a frequency region from 50 Hz to 3.4 KHz is taken into account, resulting in $k_{\text{high}}\!=\!\left \lfloor \!\frac{K}{8000}\!\cdot\!3400 \text{Hz} \!\right \rfloor\!=\!217$ and $k_{\text{low}}\!=\!\left \lfloor \!\frac{K}{8000}\!\cdot\!50 \text{Hz} \!\right \rfloor\!=\!3$ in equation (\ref{equ_lsd}) with the $512$-point FFT.   

The results are shown in Tab. \ref{tab_prelim}, in which we can see that the performance of the CNN in our proposed approach is mainly depending on the kernel length $N$, and only weakly on the choice of the activation function and the number of feature maps $F$. 
\textcolor{black}{
	Note that only a small fraction of the actually used $(N,F)$ search space is shown in Tab.~\ref{tab_prelim}: The dependence on $F$ regarding to the optimum is rather flat, however, for much smaller values of $F$ the performance deteriorates significantly.
}

As a result, the CNN topology with the minimum mean LSD value of 8.29 dB recommends the choices $F_{\text{opt}}\!=\!22$, $N_{\text{opt}}\!=\!6$, and the leaky ReLU activation function. It is interesting to know that the legacy G.711 has a mean LSD being 16.15 dB which is almost halved by applying this optimal topology. Note that $F_{\text{opt}}$ and $N_{\text{opt}}$ selected from the above preliminary experiment are specific to the framework structure \uppercase\expandafter{\romannumeral3} with the length $L\!=\!32$ of the CNN input vector. 
\textcolor{black}{In order to obtain also reasonable parameter settings for the other  framework structures, we note that the} length $L$ changes for the various postprocessing approaches with $L\!=\!\left | \mathcal{M}_{\text{env}} \right |\!=\!6.25\%\cdot K$ in the cepstral domain approaches, and $L\!=\!80$ for narrowband codecs and $L\!=\!160$ for wideband codecs in the time domain approach. \textcolor{black}{Note that for simplicity of presentation,} whenever $L$ changes \textcolor{black}{with a certain framework structure (time domain, cepstral domain \uppercase\expandafter{\romannumeral1}--\uppercase\expandafter{\romannumeral6})}, the value of $F_{\text{opt}}$ and $N_{\text{opt}}$ are \textcolor{black}{simply increased or decreased} proportionally at the same time.

\begin{table*}[!htb] 
	\begin{center}
		\footnotesize
			{\color{black}\begin{tabular}{m{0.001cm}<{\centering} m{1.8cm}<{\centering} m{1.6cm}<{\centering} | m{1.2cm}<{\centering}  m{0.6cm}<{\centering}  m{0.7cm}<{\centering}  m{0.7cm}<{\centering} m{1.1cm}<{\centering} | m{1.2cm}<{\centering}  m{0.6cm}<{\centering}  m{0.7cm}<{\centering} m{0.7cm}<{\centering} m{1.1cm}<{\centering} m{0.001cm}}
			\cline{1-13}
			& & & \multicolumn{5}{c|}{American English} & \multicolumn{5}{c}{German}& \\[3pt]
			& & & \multicolumn{2}{c}{G.711 A-law} & \multirow{2}{*}{G.726} & \multirow{2}{*}{G.722} & \multirow{2}{*}{AMR-WB} & \multicolumn{2}{c}{G.711 A-law} & \multirow{2}{*}{G.726} & \multirow{2}{*}{G.722} & \multirow{2}{*}{AMR-WB}&\\
			
			& & & no Constr. & Constr. & & & & no Constr. & Constr. & & & & \\
			\cline{1-13}
			\multicolumn{2}{c}{\multirow{2}{*}{Legacy Codec}} & \multirow{2}{*}{MOS-LQO} & \multicolumn{2}{c}{\multirow{2}{*}{\,\,\,\,\,\,\,4.29}} & \multirow{2}{*}{4.03} & \multirow{2}{*}{3.78} & \multirow{2}{*}{3.64} & \multicolumn{2}{c}{\multirow{2}{*}{\,\,\,\,\,\,\,4.14}} & \multirow{2}{*}{4.03} & \multirow{2}{*}{3.71} & \multirow{2}{*}{3.65} &\\
			& & & & & & & & & & & & &\\
			\cline{1-13}
			\multicolumn{2}{c}{\multirow{2}{*}{Postfilter~\cite{ITUTG711tollbox}}} & MOS-LQO &    \multicolumn{2}{c}{\,\,\,\,\,\,\,4.46} & \multirow{2}{*}{-} & \multirow{2}{*}{-} & \multirow{2}{*}{-} & \multicolumn{2}{c}{\,\,\,\,\,\,\,4.33} & \multirow{2}{*}{-} & \multirow{2}{*}{-} & \multirow{2}{*}{-} &\\[3pt]
			& & $\Delta$MOS-LQO & \multicolumn{2}{c}{\,\,\,\,\,\,\,0.17} &   &   &   & \multicolumn{2}{c}{\,\,\,\,\,\,\,0.19} &   &   &   &\\[3pt]
			\cline{1-13}
			
			\multicolumn{2}{c}{Structure \uppercase\expandafter{\romannumeral3}} & MOS-LQO & \bf 4.49 & \bf 4.40 & \bf 4.31 & \bf 4.73 & \bf 3.97 & \bf 4.45 & \bf 4.27  & \bf 4.38 & \bf 4.56 & \bf 3.93 &\\[3pt]
			\multicolumn{2}{c}{Cepstral Domain} & $\Delta$MOS-LQO & \bf 0.20 & \bf 0.11  & \bf 0.28 & \bf 0.95 & \bf 0.33 & \bf 0.31 & \bf 0.13 & \bf 0.35 & \bf 0.85 & \bf 0.28 &\\[3pt]
			\cline{1-13}
			
		\end{tabular}}
	\end{center}
	\textcolor{black}{\caption{MOS-LQO (\textbf{POLQA}) for legacy codecs and codecs with ITU-T postfilter~\cite{ITUTG711tollbox} and the structure \uppercase\expandafter{\romannumeral3} cepstral domain postprocessor. The best results in each column are written in \textbf{bold face}. Compare to the respective (WB-)PESQ results in Tab.~\ref{tab_dif_codecs}.}}
	\label{tab_polqa_dif_codecs}
\end{table*}

\begin{table}[]
	\begin{center}
		\footnotesize
		\begin{tabular}{m{0.001cm}<{\centering} m{2cm}<{\centering} | m{1.18cm}<{\centering}  m{0.8cm}<{\centering} | m{1.18cm}<{\centering}  m{0.3cm}<{\centering} m{0.01cm}}
			\cline{1-7}
			& & \multicolumn{2}{c|}{American English} & \multicolumn{2}{c}{\ \ \ \ \ German} & \\[4pt]
			& & no Constr. & Constr. & no Constr. & Constr. & \\
			\cline{1-7}
			& Legacy Codec  & \multicolumn{2}{c|}{\,\,\,\,37.12} & \multicolumn{2}{c}{\ \ \ \ \ \ 37.11} &\\[4pt]
			\cline{1-7}
			& Postfilter~\cite{ITUTG711tollbox} &    \multicolumn{2}{c|}{\,\,\,\,17.27} & \multicolumn{2}{c}{\ \ \ \ \ \ 15.65} & \\[4pt]
			\cline{1-7}
			& Time Domain & \bf 38.07 & \bf 38.08 & \bf 38.15 & \bf 38.15 &\\[4pt]
			\cline{1-7}
			& \uppercase\expandafter{\romannumeral1} & 29.37 & 29.98  & 29.42 & 29.99 &\\[4pt]
			\cline{3-7}
			\multirow{2}{*}{\rotatebox{90}{Cepstral Domain}} & \uppercase\expandafter{\romannumeral2} & 21.55 & 29.93  & 21.80 & 29.94 &\\[4pt]
			\cline{3-7}
			& \uppercase\expandafter{\romannumeral3} & 25.85 & 29.96 & 26.23 & 29.97 &\\[4pt]
			\cline{3-7}
			& \uppercase\expandafter{\romannumeral4} & 29.36 & 29.98 & 29.42 & 29.98 &\\[4pt]
			\cline{3-7}
			& \uppercase\expandafter{\romannumeral5} & 26.67 & 29.97 & 26.91  & 29.98 &\\[4pt]
			\cline{3-7}
			& \uppercase\expandafter{\romannumeral6} & 23.75 & 29.95 & 24.33 & 29.95 &\\[4pt]
			\cline{1-7}
		\end{tabular}
	\end{center}
	\caption{The \textbf{$\text{SSDR}_{\text{seg}}$ [dB]} values for the G.711 legacy codec and G.711 codec with various postprocessors. The best approach is written in \textbf{bold face}.}
	\label{tab_dif_codecs_ssdr}
	\vspace{-6pt}
\end{table}

\textcolor{black}{
	Now, as we have fixed the number of trainable parameters, we briefly want to check whether a straight-forward fully-connected neural network (FCNN) performs equally well. As shown in Tab.\;\uppercase\expandafter{\romannumeral3}, we simulated four different FCNN topologies without dropout or a dropout rate $r\!=\!0.1$ for both the time domain approach and the cepstral domain approach of structure \uppercase\expandafter{\romannumeral3}, while keeping the same number of input nodes ($L\!=\!80$ for the time domain approach and $L\!=\!32$ for the cepstral domain approach). The number of trainable parameters is about the same as (or a bit higher than) the optimal CNN topology with ($F_{\text{opt}},N_{\text{opt}}$). It can be seen that the optimal CNN topology achieves the best LSD performance compared to all listed FCNN structures for both the time domain approach and the cepstral domain approach. 
	Accordingly, in the following we stick to the CNN topology as it seems to be an advantageous choice. 
}

\subsection{Major Instrumental Experiments} \label{subsec_simu_dis}
In this subsection, the experiments of the proposed postprocessing approaches for the various codecs in different conditions are implemented and evaluated instrumentally following the test processing in Fig. \ref{figures_process_plan2}. 

\subsubsection{Clean Condition}
\label{clean_conditions}
A comprehensive evaluation of all the proposed postprocessors is conducted for four different codecs in both American English and German language, in which legacy codecs and the postfilter for G.711 serve as baselines. PESQ results are shown in Tab. \ref{tab_dif_codecs} with $\Delta$MOS-LQO being the MOS-LQO difference between the postfilter or the postprocessor and the respective legacy codec. We find that most of our proposed postprocessors perform better than the respective legacy codecs. For G.711 our proposed postprocessors in most cases show better performance when no quantization constraint is performed. Comparing the various proposed postprocessors with no quantization constraint, the time domain postprocessor and the cepstral domain postprocessors with structures \uppercase\expandafter{\romannumeral2}, \uppercase\expandafter{\romannumeral3}, \uppercase\expandafter{\romannumeral5}, and \uppercase\expandafter{\romannumeral6} (the ones with delay, see Tab.~\ref{table_structures}) show \emph{better performance than all legacy codecs} and they \emph{all perform better than or equal to the G.711 postfilter}~\cite{ITUTG711tollbox} for both languages (only the time domain postprocessor has the same MOS-LQO as the postfilter for American English). The cepstral domain postprocessor with structure \uppercase\expandafter{\romannumeral6} performs best for both languages and for all codecs, exceeding the legacy codecs on average over both languages by 0.25 MOS points for G.711, 0.3 MOS points for G.726, and 0.26 MOS points for AMR-WB. Note that structure \uppercase\expandafter{\romannumeral6} exceeds the G.722 legacy codec by an impressive 0.82 MOS points\textcolor{black}{, where roughly 0.3 MOS points can be dedicated to the rather simple suppression of frequencies beyond 7 kHz, and the major rest can be dedicated indeed to the improvement of the early cepstral coefficients (details are given in the Appendix)}. 

\begin{table}[]
	\begin{center}
		\footnotesize
		{\color{black}\begin{tabular}{m{0.001cm}<{\centering} m{2cm}<{\centering} | m{1.18cm}<{\centering}  m{0.8cm}<{\centering} | m{1.18cm}<{\centering}  m{0.3cm}<{\centering} m{0.01cm}}
				\cline{1-7}
				& & \multicolumn{2}{c|}{American English} & \multicolumn{2}{c}{\ \ \ \ \ German} & \\[4pt]
				& & no Constr. & Constr. & no Constr. & Constr. & \\
				\cline{1-7}
				& Legacy Codec  & \multicolumn{2}{c|}{\,\,\,\,37.30} & \multicolumn{2}{c}{\ \ \ \ \ \ 37.35} &\\[4pt]
				\cline{1-7}
				& Postfilter~\cite{ITUTG711tollbox} &    \multicolumn{2}{c|}{\,\,\,\,17.79} & \multicolumn{2}{c}{\ \ \ \ \ \ 15.70} & \\[4pt]
				\cline{1-7}
				& Time Domain & \bf 38.33 & \bf 38.34 & \bf 38.49 & \bf 38.50 &\\[4pt]
				\cline{1-7}
				& \uppercase\expandafter{\romannumeral1} & 29.98 & 34.15  & 30.05 & 34.17 &\\[4pt]
				\cline{3-7}
				\multirow{2}{*}{\rotatebox{90}{Cepstral Domain}} & \uppercase\expandafter{\romannumeral2} & 19.94 & 32.36  & 20.54 & 32.35 &\\[4pt]
				\cline{3-7}
				& \uppercase\expandafter{\romannumeral3} & 24.58 & 33.04 & 25.14 & 33.03 &\\[4pt]
				\cline{3-7}
				& \uppercase\expandafter{\romannumeral4} & 29.93 & 34.11 & 30.11 & 34.15 &\\[4pt]
				\cline{3-7}
				& \uppercase\expandafter{\romannumeral5} & 25.39 & 32.97 & 25.83  & 32.93 &\\[4pt]
				\cline{3-7}
				& \uppercase\expandafter{\romannumeral6} & 21.80 & 32.66 & 23.26 & 32.71 &\\[4pt]
				\cline{1-7}
			\end{tabular}}
		\end{center}
		\vspace{-4pt}
		\textcolor{black}{\caption{The \textbf{$\text{SSDR}$ [dB]} values for the G.711 legacy codec and G.711 codec with various postprocessors. The best approach is written in \textbf{bold face}.}}
		\label{tab_snr}
		\vspace{-8.5pt}
\end{table}

\textcolor{black}{
	For a limited set of conditions in Tab.\;\ref{tab_dif_codecs}, we provide also POLQA~\cite{ITUT_polqa_2018} results in Tab.\;\uppercase\expandafter{\romannumeral5}. Note that very similar improvements of our postprocessor (structure \uppercase\expandafter{\romannumeral3}) w.r.t.\ all legacy codecs in both languages can be seen, with the AMR-WB postprocessor performing even better in POLQA than in WB-PESQ. However, since simulation of PESQ was much easier to perform due to the availability of a batch mode to us, the remainder of our work uses PESQ and WB-PESQ.}

\textcolor{black}{
	In order to obtain a better understanding of how the coded speech signal is enhanced by the cepstral domain approach, spectral and cepstral analysis examples of the enhanced speech, along with the coded and reference speech, are presented for interested readers in the Appendix. 
	}

Since the algorithmic delay might be critical in practical applications, we see that the zero-latency time domain postprocessors can improve the speech quality for all listed codecs in both languages.
For cepstral domain postprocessors, the zero-latency structures \uppercase\expandafter{\romannumeral1} and \uppercase\expandafter{\romannumeral4} still can consistently improve speech quality of G.726 and particularly of G.722. Since G.711 and AMR-WB ask for some delay in the postprocessor, a good compromise for these codecs would be the structure \uppercase\expandafter{\romannumeral3}, providing second ranked speech quality in both languages. At the cost of only 10 ms algorithmic delay, structure \uppercase\expandafter{\romannumeral3} exceeds the legacy codecs on average over both languages by 0.25 MOS points for G.711, 0.25 MOS points for G.726, 0.81 MOS points for G.722, and 0.2 MOS points for AMR-WB.

\textcolor{black}{
	To further illustrate the influence of the additional delay on the performance improvement, we compare in Fig.\ 9 the MOS-LQO of the postfilter and the postprocessors both in time domain and cepstral domain for G.711, sorted by the additional delay. The MOS-LQO is an average of American English and German. For the proposed postprocessors in the cepstral domain, it becomes obvious that the performance improvement grows with the increase of the additional delay, as the model topology is exactly the same (except for structure \uppercase\expandafter{\romannumeral2}\footnote{\textcolor{black}{This is because structure \uppercase\expandafter{\romannumeral2} has a different topology in terms of the number of input nodes $L$, feature maps $F$, and kernel size $N$ (see Tab.\;\ref{tab_relative_complex}).}}, see Tab.\;\ref{tab_relative_complex}). With longer additional delay for the postfilter~\cite{ITUTG711tollbox}, it may also achieve some further performance gains. However, our proposed zero-latency postprocessor in the time domain already shows superior performance compared to the ITU-T postfilter with 2 ms additional delay. 
}

Comparing the bold face (i.e., top-two) results in Tab.~\ref{tab_dif_codecs}, we see that there is hardly a language dependency in the rank order of the best approaches.

\begin{figure}[tp]
	
	\psfrag{4.1}[cr][cr]{\footnotesize $4.1$}
	\psfrag{4.15}[cr][cr]{\footnotesize $4.15$}
	\psfrag{4.2}[cr][cr]{\footnotesize $4.2$}
	\psfrag{4.25}[cr][cr]{\footnotesize $4.25$}
	\psfrag{4.3}[cr][cr]{\footnotesize $4.3$}
	\psfrag{4.35}[cr][cr]{\footnotesize $4.35$}
	\psfrag{4.4}[cr][cr]{\footnotesize $4.4$}
	\psfrag{4.45}[cr][cr]{\footnotesize $4.45$}
	\psfrag{4.5}[cr][cr]{\footnotesize $4.5$}
	
	\psfrag{0}[tc][tc]{\footnotesize $0$}
	\psfrag{2}[tc][tc]{\footnotesize $2$}
	\psfrag{5}[tc][tc]{\footnotesize $5$}
	\psfrag{10}[tc][tc]{\footnotesize $10$}
	\psfrag{16}[tc][tc]{\footnotesize $16$}
	
	\psfrag{PESQ}[cb][ct]{\footnotesize MOS-LQO}
	\psfrag{AddDel}[ct][cb]{\footnotesize Additional Delay (ms)}
	\psfrag{Legacccyg7111}[cc][cc]{\footnotesize Legacy G.711}
	
	\psfrag{a}[cc][cc]{\footnotesize \uppercase\expandafter{\romannumeral1}}
	\psfrag{b}[cc][cc]{\footnotesize \uppercase\expandafter{\romannumeral4}}
	\psfrag{c}[cc][cc]{\footnotesize Domain}
	\psfrag{z}[tc][bc]{\footnotesize Time}
	\psfrag{w}[tc][bc]{\footnotesize ITU-T}
	\psfrag{y}[tc][bc]{\footnotesize Postfilter}
	\psfrag{d}[tc][bc]{\footnotesize \cite{ITUTG711tollbox}}
	\psfrag{e}[cc][cc]{\footnotesize \uppercase\expandafter{\romannumeral5}}
	\psfrag{f}[cc][cc]{\footnotesize \uppercase\expandafter{\romannumeral3}}
	\psfrag{g}[cc][cc]{\footnotesize \uppercase\expandafter{\romannumeral6}}
	\psfrag{x}[cc][cc]{\footnotesize \uppercase\expandafter{\romannumeral2}}
	
	\centering
	\includegraphics[height=\columnwidth,angle=270]{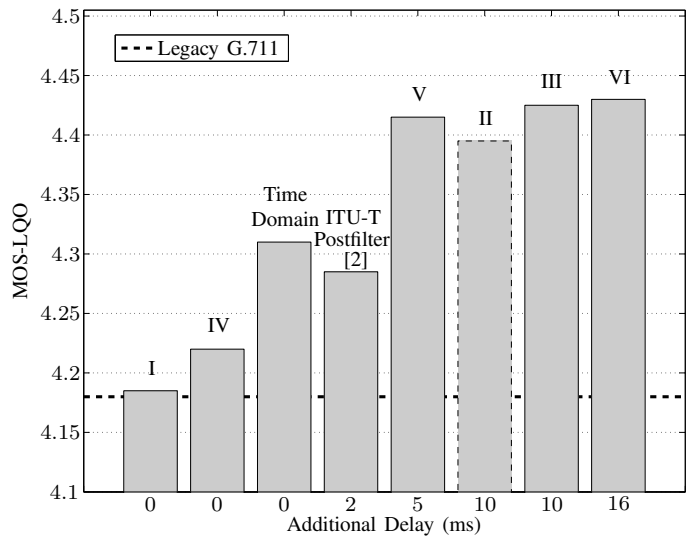}  
	\vspace{1pt} 
	\textcolor{black}{\caption{The MOS-LQO (\textbf{PESQ}) of various postprocessors for G.711 with different amounts of additional delay. Note that structure \uppercase\expandafter{\romannumeral2} has a different model topology than the other cepstral domain structures \uppercase\expandafter{\romannumeral1}, \uppercase\expandafter{\romannumeral3}-\uppercase\expandafter{\romannumeral6}; see Tab.\;\ref{tab_relative_complex}.}}
	\label{fig_delays_g711}
	\vspace*{-2pt}
\end{figure}

To intuitively show the potential of the postprocessor with structure \uppercase\expandafter{\romannumeral3} we performed a comparison to different modes (i.e., bitrates) for the AMR-WB codec in Fig.~\ref{fig_amrwb_opp}. One can easily see that the MOS-LQO of the postprocessor after the AMR-WB codec at 12.65 kbps for both American English and German exceeds the legacy AMR-WB at 15.85 kbps and it even approaches a comparable quality for German at 18.25 kbps. Therefore, the postprocessor with structure \uppercase\expandafter{\romannumeral3} shows its ability to significantly improve the speech quality during transmission with a relative low bitrate towards a much higher bitrate transmission.

In order to see the waveform distortion of speech after the postprocessing, $\text{SSDR}_{\text{seg}}$ measure (\ref{equ_ssdr}) for G.711 A-law in American English and German is shown in Tab.~\ref{tab_dif_codecs_ssdr}. It is straightforward that the legacy G.711 already achieves a relatively high $\text{SSDR}_{\text{seg}}$, with 37.12 dB for American English and 37.11 dB for German, since it is a high bitrate waveform coding. 
For the time domain postprocessor, it achieves even higher $\text{SSDR}_{\text{seg}}$ which is the best performance among all the proposed postprocessors, since it focuses on the waveform domain. 
All proposed postprocessors with quantization constraint show equal or better $\text{SSDR}_{\text{seg}}$ than without it, but it brings no positive effect to speech quality for the proposed postprocessors in terms of MOS-LQO (see Tab.~\ref{tab_dif_codecs}). 
For the postprocessor with structure \uppercase\expandafter{\romannumeral6}, which achieves the best speech quality (see Tab.~\ref{tab_dif_codecs}), a mean $\text{SSDR}_{\text{seg}}$ of only 24.04 dB over both languages is measured. Comparing $\text{SSDR}_{\text{seg}}$ and MOS-LQO, we once again see that waveform similarity and speech quality are not necessarily positively correlated, in this case also questioning the quantization constraint.

\textcolor{black}{Evaluating the global SSDR measure in Tab.\;\uppercase\expandafter{\romannumeral7}, it turns out that the rank order of approaches is very similar to the $\text{SSDR}_{\text{seg}}$ in Tab.~\ref{tab_dif_codecs_ssdr}: The proposed time domain approach is the best, followed by the G.711 legacy codec, the cepstral domain approaches, and finally the ITU-T postfilter~\cite{ITUTG711tollbox}. Interestingly, the advantage of using the constraint is higher with the SSDR measure, which might be due to some very slight residual noise for the cepstral domain approaches in speech pauses; an effect that has been disregarded in $\text{SSDR}_{\text{seg}}$ through the inherent voice activity detection in (\ref{equ_ssdr}), and which will motivate some small extra processing in Section \ref{subsec_subjective_exp}. }

\begin{figure}[tp]
	
	\psfrag{0}[br][cc]{\footnotesize $10$}
	\psfrag{3.5}[cr][cr]{\footnotesize $3.5$}
	\psfrag{3.55}[cr][cr]{\footnotesize $3.55$}
	\psfrag{3.6}[cr][cr]{\footnotesize $3.6$}
	\psfrag{3.65}[cr][cr]{\footnotesize $3.65$}
	\psfrag{3.7}[cr][cr]{\footnotesize $3.7$}
	\psfrag{3.75}[cr][cr]{\footnotesize $3.75$}
	\psfrag{3.8}[cr][cr]{\footnotesize $3.8$}
	\psfrag{3.85}[cr][cr]{\footnotesize $3.85$}
	\psfrag{3.9}[cr][cr]{\footnotesize $3.9$}
	\psfrag{3.95}[cr][cr]{\footnotesize $3.95$}
	\psfrag{4}[cr][cr]{\footnotesize $4$}	
	
	\psfrag{12}[tc][tc]{\footnotesize $12.65$}
	\psfrag{14}[tc][tc]{\footnotesize $14.25$}
	\psfrag{15}[tc][tc]{\footnotesize $15.85$}
	\psfrag{18}[tc][tc]{\footnotesize $18.25$}
	\psfrag{19}[tc][tc]{\footnotesize $19.85$}
	\psfrag{23}[tr][tr]{\footnotesize $23.05$}
	\psfrag{24}[tc][tr]{\footnotesize $23.85$}
	
	\psfrag{G}[bl][tl]{\footnotesize $G_p \left [\text{dB}\right ]$}
	\psfrag{N}[tc][bl]{\footnotesize $N_p$}
	
	\psfrag{Postprocessed}[cl][cl]{\footnotesize $\text{Postprocessor after AMR-WB at } 12.65 \text{ kbps} $}
	\psfrag{with}[cl][cl]{\footnotesize $\text{with structure \uppercase\expandafter{\romannumeral3}}$}
	\psfrag{B}[cl][bl]{\footnotesize $\text{AMR-WB}$}
	\psfrag{C}[cl][bl]{\footnotesize $\text{at various bitrates}$}
	
	\psfrag{Americanenglishameri}[tl][tl]{\footnotesize $\text{American English}$}
	\psfrag{Germangerm}[tl][tl]{\footnotesize $\text{German}$}
	
	\psfrag{A}[cc][tc]{\footnotesize $\text{Bitrate}\left [ \text{kbps} \right ]$}
	\psfrag{M}[cc][cc]{\footnotesize $\text{MOS-LQO}$} 
	
	\centering
	\vspace*{-1pt}
	\includegraphics[height=1.01\columnwidth,angle=270]{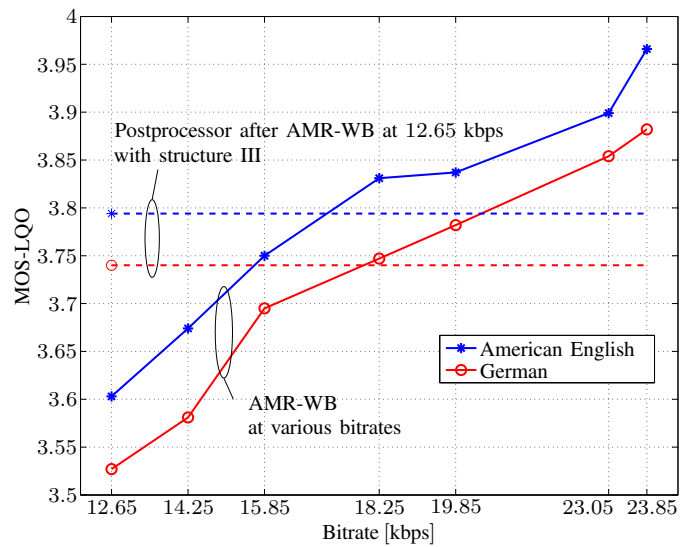}   
	\caption{MOS-LQO (\textbf{WB-PESQ}) points of the test speech utterances for the legacy AMR-WB at various bitrates (solid curves from 12.65 to 23.85 kbps) as well as for our postprocessor with structure \uppercase\expandafter{\romannumeral3} at 12.65 kbps (dashed lines). Results are shown for American English ($\color{blue} \bm{\ast}$) and German ($\color{red} \bm{\circ}$).}
	\label{fig_amrwb_opp}
	\vspace*{-5pt}
\end{figure}

\begin{table*}[!htb] 
	\begin{center}
		\footnotesize
		\begin{tabular}{ m{2.3cm}<{\centering}  m{1.7cm}<{\centering} |   m{1.15cm}<{\centering}  m{1.15cm}<{\centering} m{1.15cm}<{\centering}  |  m{1.2cm}<{\centering} m{1.2cm}<{\centering} m{1.29cm}<{\centering} m{0.001cm}}
			\cline{1-8}
			& & \multicolumn{3}{c|}{Narrowband Tandeming} & \multicolumn{3}{c}{Wideband Tandeming}& \\[3pt]
			
			& &  $\mu$-law\,+ A-law & G.726\,+ A-law & AMR\,+ A-law &  G.711.1  (A-law)\,+ AMR-WB & G.722\,+ AMR-WB & EVS-WB\,+ AMR-WB & \\[23pt]
			
			\cline{1-8}
			\multirow{2}{*}{Legacy Codec} & \multirow{2}{*}{MOS-LQO} & \multirow{2}{*}{4.18} & \multirow{2}{*}{3.96} & \multirow{2}{*}{4.01} & \multirow{2}{*}{3.37} & \multirow{2}{*}{3.34} & \multirow{2}{*}{3.28}  &\\
			& & & & & & & &  \\
			\cline{1-8}
			\multirow{2}{*}{Postfilter~\cite{ITUTG711tollbox}} & MOS-LQO & 4.20 & 4.01 & 4.07 & \multirow{2}{*}{-} & \multirow{2}{*}{-} & \multirow{2}{*}{-} &  \\[3pt]
			& $\Delta$MOS-LQO & 0.02 & 0.05 & 0.06 &   &   &   &  \\[3pt]
			\cline{1-8}
			\multirow{2}{*}{Time Domain} & MOS-LQO & 4.28 & 4.03 & 4.09 & 3.39 & 3.49 & 3.29 &  \\[3pt]
			& $\Delta$MOS-LQO & 0.10 & 0.07 & 0.08 & 0.02 & 0.15 & 0.01 & \\[3pt]
			\cline{1-8}
			Structure \uppercase\expandafter{\romannumeral3} & MOS-LQO & 4.38 & 4.13 & 4.12 & 3.70  & 3.71 & 3.48 &  \\[3pt]
			Cepstral Domain& $\Delta$MOS-LQO & 0.20 & 0.17 & 0.11 & 0.33 & 0.37 & 0.20 &  \\[3pt]
			\cline{1-8}
			Structure \uppercase\expandafter{\romannumeral6} & MOS-LQO & \bf 4.41 & \bf 4.18 & \bf 4.13 & \bf 3.78  & \bf 3.75 & \bf 3.53 &  \\[3pt]
			Cepstral Domain& $\Delta$MOS-LQO & \bf 0.23 & \bf 0.22 & \bf 0.12 & \bf 0.41 & \bf 0.41 & \bf 0.25 &  \\[3pt]
			\cline{1-8}
		\end{tabular}
	\end{center}
	\caption{MOS-LQO (\textbf{PESQ} and \textbf{WB-PESQ}) for legacy codecs and codecs with different postprocessors in \textbf{tandeming conditions}. \protect\\ The results of the best approach is written in \textbf{bold face}.}
	\label{tab_tandem_nb_wb}
	\vspace{-3pt}
\end{table*}

\begin{table*}[!htb] 
	\begin{center}
		\footnotesize
		\begin{tabular}{ m{2.3cm}<{\centering}  m{1.7cm}<{\centering} |   m{1cm}<{\centering}  m{1cm}<{\centering} m{1cm}<{\centering}  m{1cm}<{\centering} | m{1cm}<{\centering}  m{1cm}<{\centering} m{1cm}<{\centering} m{1cm}<{\centering} m{0.001cm}}
			\cline{1-10}
			& & \multicolumn{4}{c|}{G.711} & \multicolumn{4}{c}{AMR-WB}& \\[3pt]
			& & \multicolumn{2}{c}{Random} & \multicolumn{2}{c|}{Burst} & \multicolumn{2}{c}{Random} & \multicolumn{2}{c}{Burst} & \\
			
			& & 3$\%$ & 6$\%$ & 3$\%$ & 6$\%$ & 3$\%$ & 6$\%$ & 3$\%$ & 6$\%$ & \\[3pt]
			
			\cline{1-10}
			\multirow{2}{*}{Legacy Codec} & \multirow{2}{*}{MOS-LQO} & \multirow{2}{*}{3.67} & \multirow{2}{*}{3.31} & \multirow{2}{*}{3.60} & \multirow{2}{*}{3.07} & \multirow{2}{*}{2.75} & \multirow{2}{*}{2.30} & \multirow{2}{*}{2.80} & \multirow{2}{*}{2.39} &\\
			& & & & & & & & & & \\
			\cline{1-10}
			\multirow{2}{*}{Postfilter~\cite{ITUTG711tollbox}} & MOS-LQO & 3.71 & 3.34 & 3.66 & 3.12 & \multirow{2}{*}{-} & \multirow{2}{*}{-} & \multirow{2}{*}{-} & \multirow{2}{*}{-} &\\[3pt]
			& $\Delta$MOS-LQO & 0.04 & 0.03 & 0.06 & 0.05 &   &   &   &   &\\[3pt]
			\cline{1-10}
			\multirow{2}{*}{Time Domain} & MOS-LQO & 3.71 & 3.35 & 3.67 & 3.12 & 2.78 & 2.32 & 2.83 & 2.41 &\\[3pt]
			& $\Delta$MOS-LQO & 0.04 & 0.04 & 0.07 & 0.05 & 0.03 & 0.02 & 0.03 & 0.02 &\\[3pt]
			\cline{1-10}
			Structure \uppercase\expandafter{\romannumeral3} & MOS-LQO & 3.74 & 3.37 & \bf 3.76  & \bf 3.19  & 2.94 & 2.44 & 2.99 & 2.54 &\\[3pt]
			Cepstral Domain& $\Delta$MOS-LQO & 0.07 & 0.06 & \bf 0.16 &  \bf 0.12 & 0.19 & 0.14 & 0.19 & 0.15 &\\[3pt]
			\cline{1-10}
			Structure \uppercase\expandafter{\romannumeral6} & MOS-LQO & \bf 3.76 & \bf 3.41 &  3.73 & 3.16   & \bf 3.03  & \bf 2.51 & \bf 3.09 & \bf 2.62 &  \\[3pt]
			Cepstral Domain& $\Delta$MOS-LQO & \bf 0.09 & \bf 0.10 &  0.13 & 0.09  & \bf 0.28 & \bf 0.21 & \bf 0.29 & \bf 0.23 &  \\[3pt]
			\cline{1-10}
		\end{tabular}
	\end{center}
	\caption{MOS-LQO (\textbf{PESQ} and \textbf{WB-PESQ}) for G.711 and AMR-WB legacy codecs and codecs with different postprocessors in \textbf{error-prone transmission conditions}.  The results of the best approach is written in \textbf{bold face}.}
	\label{tab_error_channel}
	\vspace{-3pt}
\end{table*}

\definecolor{lightgray}{gray}{0.9}
\begin{table*}[!htb] 
	\begin{center}
		\footnotesize
		\setlength\tabcolsep{4.5pt} 
		\begin{tabular}{ m{1.93cm}<{\centering}  m{1.53cm}<{\centering} |   m{0.5cm}<{\centering} m{0.5cm}<{\centering} m{0.59cm}<{\centering} m{0.59cm}<{\centering} m{0.59cm}<{\centering} m{0.5cm}<{\centering} m{0.5cm}<{\centering} m{0.5cm}<{\centering} | m{0.59cm}<{\centering} m{0.59cm}<{\centering} m{0.59cm}<{\centering}  m{0.59cm}<{\centering} m{0.59cm}<{\centering} m{0.5cm}<{\centering} m{0.59cm}<{\centering} m{0.6cm}<{\centering} m{0.001cm}}
			\cline{1-18}
			& & \multicolumn{8}{c|}{G.711} & \multicolumn{8}{c}{AMR-WB}& \\[3pt]
			
			& & \multicolumn{2}{c}{Cafeteria} & \multicolumn{2}{c}{Car} & \multicolumn{2}{c}{Road} &\multirow{2}{*}{Mean} & \multirow{2}{*}{Clean} & \multicolumn{2}{c}{Cafeteria} & \multicolumn{2}{c}{Car} & \multicolumn{2}{c}{Road} &\multirow{2}{*}{Mean} & \multirow{2}{*}{Clean} &  \\[3pt]
			SNR [dB] & & 15 & 20 & 15 & 20 & 15 & 20 & & & 15 & 20 & 15 & 20 & 15 & 20 & & &\\[3pt]
			\cline{1-18}
			\multirow{2}{*}{Legacy Codec} & \multirow{2}{*}{MOS-LQO} & \multirow{2}{*}{2.29} & \multirow{2}{*}{2.67} & \multirow{2}{*}{2.40} & \multirow{2}{*}{2.75} & \multirow{2}{*}{2.06} & \multirow{2}{*}{2.43} & \multirow{2}{*}{2.43} & \multirow{2}{*}{4.21} & \multirow{2}{*}{1.69} & \multirow{2}{*}{2.10} & \multirow{2}{*}{2.12} & \multirow{2}{*}{2.52} & \multirow{2}{*}{1.59} & \multirow{2}{*}{1.99} & \multirow{2}{*}{2.00} & \multirow{2}{*}{3.60} &\\
			& & & & & & & & & & & & & & & & \\
			\cline{1-18}
			\multirow{2}{*}{Postfilter~\cite{ITUTG711tollbox}} & MOS-LQO & \bf 2.31 & 2.70 & \bf 2.41 & \bf 2.76 & 2.07 & 2.45 & 2.45 & 4.32 & \multirow{2}{*}{-} &\multirow{2}{*}{-} & \multirow{2}{*}{-} &\multirow{2}{*}{-} & \multirow{2}{*}{-} &\multirow{2}{*}{-} &\multirow{2}{*}{-}&\multirow{2}{*}{-} & \\[3pt]
			& $\Delta$MOS-LQO & \bf 0.02 & 0.03 & \bf 0.01 & \bf 0.01 & 0.01 & 0.02 & 0.02 & 0.11 & &  & &  & &  & & &  \\[3pt]
			\cline{1-18}
			\multirow{2}{*}{Time Domain} & MOS-LQO & \bf 2.31 & 2.69 & \bf 2.41 & \bf 2.76 & 2.06 & 2.45 & 2.45 & 4.32 & 1.73 & 1.74 & 1.98 & 2.39 &  1.63 & 2.05 &1.92 & 3.61 &  \\[3pt]
			& $\Delta$MOS-LQO & \bf 0.02 & 0.02 & \bf 0.01 & \bf 0.01  & 0 & 0.02 & 0.02 & 0.11 & 0.04 & -0.36 & -0.14 & -0.13 & 0.04 & 0.06 &-0.08 & 0.01 & \\[3pt]
			\cline{1-18}
			Structure \uppercase\expandafter{\romannumeral3} & MOS-LQO & 2.29 & 2.68 & 2.39 & 2.75 & 2.05 & 2.43 & 2.43 & 4.43 & 1.68 &  2.13 &  2.25 & 2.68 & 1.56 & 1.99 & 2.05 &3.79 &  \\[3pt]
			Cepstral Domain& $\Delta$MOS-LQO & 0 & 0.01 & -0.01 & 0 & -0.01 & 0 & 0 & 0.22 & -0.01 &  0.03 &  0.13 & 0.16 & -0.03 & 0 &0.05 & 0.19 &  \\[3pt]
			
			\hhline{------------------}
			\cellcolor[gray]{0.9} Structure \uppercase\expandafter{\romannumeral3} &\cellcolor[gray]{0.9} MOS-LQO &\cellcolor[gray]{0.9} \bf 2.31 
			&\cellcolor[gray]{0.9} \bf 2.73 &\cellcolor[gray]{0.9} 2.38 
			&\cellcolor[gray]{0.9} 2.70 &\cellcolor[gray]{0.9} \bf 2.19 
			&\cellcolor[gray]{0.9} \bf 2.60 &\cellcolor[gray]{0.9} \bf 2.49 
			&\cellcolor[gray]{0.9} 4.32 &\cellcolor[gray]{0.9} \bf 1.75 
			&\cellcolor[gray]{0.9} \bf 2.19 &\cellcolor[gray]{0.9} 2.24 
			&\cellcolor[gray]{0.9} 2.63 &\cellcolor[gray]{0.9} \bf 1.80 
			&\cellcolor[gray]{0.9} \bf 2.22 &\cellcolor[gray]{0.9} \bf 2.14
			&\cellcolor[gray]{0.9} 3.66 &  \\[3pt]
			\cellcolor[gray]{0.9} Cepstral Domain& \cellcolor[gray]{0.9} $\Delta$MOS-LQO 
			&\cellcolor[gray]{0.9} \bf 0.02 &\cellcolor[gray]{0.9} \bf 0.06
			&\cellcolor[gray]{0.9} -0.02 &\cellcolor[gray]{0.9} -0.05 
			&\cellcolor[gray]{0.9} \bf 0.13 &\cellcolor[gray]{0.9} \bf 0.17 
			&\cellcolor[gray]{0.9} \bf 0.06 &\cellcolor[gray]{0.9} 0.11 
			&\cellcolor[gray]{0.9} \bf 0.06 &\cellcolor[gray]{0.9}  \bf 0.09 &\cellcolor[gray]{0.9}  0.12 &\cellcolor[gray]{0.9} 0.11 
			&\cellcolor[gray]{0.9} \bf 0.21 &\cellcolor[gray]{0.9} \bf 0.23
			&\cellcolor[gray]{0.9} \bf 0.14 &\cellcolor[gray]{0.9} 0.06 &  \\[3pt]
			\hhline{------------------}
			Structure \uppercase\expandafter{\romannumeral6} & MOS-LQO & 2.30 & 2.69 & 2.40 & 2.75 & 2.06 & 2.45 & 2.44 & \bf 4.44 &  1.71 & 2.17 & \bf 2.29 & \bf 2.76 & 1.59  & 2.05 & 2.10 & \bf 3.85 &  \\[3pt]
			Cepstral Domain& $\Delta$MOS-LQO & 0.01 & 0.02 & 0 & 0 & 0 & 0.02 &0.01& \bf 0.23 &  0.02  & 0.07 & \bf 0.17 & \bf 0.24 & 0 & 0.06 &0.10& \bf 0.25 &  \\[3pt]
			\cline{1-18}
		\end{tabular}
	\end{center}
	\caption{MOS-LQO (\textbf{PESQ} and \textbf{WB-PESQ}) for G.711 and AMR-WB legacy codecs and codecs with different postprocessors in \textbf{noisy speech conditions}. \textcolor{black}{The results of the best approach is written in \textbf{bold face} and the model trained with 20 dB (unseen) noisy data is \colorbox{lightgray}{grey-shaded}.}}
	\label{tab_noisy}
	\vspace{-12pt}
\end{table*}

\subsubsection{Tandeming Conditions}
\label{tandeming_conditions}
In order to evaluate the performance of the proposed postprocessors in tandeming conditions, G.711 A-law and AMR-WB are selected as the last codec for narrowband and wideband, respectively, while several other codecs form some common tandeming conditions. The CNN model matches the last codec since only this codec is known at the receiving point. It is worth noting that all further experiments in this subsection are only conducted in American English. The PESQ results are shown in Tab. \ref{tab_tandem_nb_wb} and we can see the performance of our time domain postprocessor and the postprocessor in the cepstral domain with structures \uppercase\expandafter{\romannumeral3} and \uppercase\expandafter{\romannumeral6}. While in narrowband tandem conditions structure \uppercase\expandafter{\romannumeral3} achieves a MOS-LQO improvement in the range $0.11...0.20$ points (in all cases the postprocessor has been just trained for the receiving-sided A-law G.711), the structure \uppercase\expandafter{\romannumeral3} in wideband tandeming conditions improves by $0.20...0.37$ PESQ MOS points (the postprocessor has been only trained for the receiving-sided AMR-WB). Note that the G711.1 A-law\,+\,AMR-WB tandeming and the G.722\,+\,AMR-WB tandeming, followed by structure \uppercase\expandafter{\romannumeral3} both achieve around 3.7 PESQ MOS points, which is even more than only AMR-WB with 3.6 points (see Tab. \ref{tab_dif_codecs}). With the best postprocessor of structure \uppercase\expandafter{\romannumeral6} from Tab.~\ref{tab_dif_codecs}, even slightly better speech quality is achieved in all cases for the price of a large algorithmic delay. \emph{All of the postprocessors in Tab.\ \ref{tab_tandem_nb_wb} exceed the shown legacy codecs under tandeming, even if the legacy codec (G.711 A-law) is followed by the postfilter from~\cite{ITUTG711tollbox}}.

\subsubsection{Error-Prone Transmission Conditions}
\label{error_pron_conditions}
For the evaluation of the proposed postprocessors in error-prone transmission, random and burst frame losses are inserted to the bitstream of G.711 and AMR-WB with the FER being $3\%$ and $6\%$ and the PESQ results are shown in Tab.\ \ref{tab_error_channel}. It is worth noting that the error concealment measures are applied in all conditions for both codecs: the packet loss concealment for G.711 from Appendix \uppercase\expandafter{\romannumeral1}~\cite{ITU_G711_PLC} and the error concealment of erroneous or lost frames for AMR-WB from 3GPP TS 26.191~\cite{3GPP_26191}. \textcolor{black}{Note that AMR-WB in this condition requires DTX to be switched on.}
The time domain postprocessor has better or equal performance compared to the postfilter\cite{ITUTG711tollbox} for G.711 for both random and burst frame losses, and is very slightly better in the case of AMR-WB. The cepstral domain postprocessors with structures \uppercase\expandafter{\romannumeral3} and \uppercase\expandafter{\romannumeral6} both perform even better in all cases and structure \uppercase\expandafter{\romannumeral3} with less delay improves the legacy codecs by $0.06...0.16$ PESQ MOS points in narrowband frame loss and $0.14...0.19$ PESQ MOS points in wideband frame loss.
\emph{Accordingly, all of the postprocessors in Tab.\ \ref{tab_error_channel} can be advantageously employed after the legacy codecs in frame loss conditions}.

\begin{table}[]
	\begin{center}
		\footnotesize
		\begin{tabular}{m{5cm}<{\centering} | m{0.8cm}<{\centering} | m{1.5cm}<{\centering}  m{0.001cm}}
			\cline{1-3}
			
			CCR Cases & CMOS & $CI_{95}$ & \\[4pt]
			\cline{1-3}
			
			Legacy G.711 vs. \bf Direct & 1.76 & $[1.61;1.92]$ &\\[4pt]
			\cline{1-3}
			
			Postfilter~\cite{ITUTG711tollbox} vs. \bf Direct & 0.28 & $[0.13;0.43]$ &\\[4pt]
			\cline{1-3}
			
			{\bf Proposed Postprocessor} vs. Direct & -0.18 & $[\text{-}0.33;\text{-}0.02]$ & \\[4pt]
			\cline{1-3}	
			
			Legacy G.711 vs. \bf Postfilter~\cite{ITUTG711tollbox} & 1.45 &$[1.27;1.64]$  & \\[4pt]
			\cline{1-3}
			
			Legacy G.711 vs. \bf Proposed Postprocessor & 1.77 & $[1.60;1.95]$ & \\[4pt]
			
			\cline{1-3}
			
			Postfilter~\cite{ITUTG711tollbox} vs. \bf Proposed Postprocessor & 0.36 & $[0.23;0.50]$ & \\[4pt]
			\cline{1-3}
			
		\end{tabular}
	\end{center}
	\caption{\textbf{CCR subjective listening test results} with the baseline postfilter~\cite{ITUTG711tollbox}, the proposed postprocessor of structure \uppercase\expandafter{\romannumeral3}, the legacy G.711 codec and the direct condition. The winning condition is written in \textbf{bold face}.}
	\label{tab_subject}
	\vspace{-4pt}
\end{table}

\subsubsection{Noisy Speech Conditions}
\label{noisy_conditions}
In order to evaluate the performance of the proposed postprocessing approaches for noisy speech, different types of background noise are added to the speech signals at an SNR of 15 dB or 20 dB, followed by G.711 and AMR-WB. The PESQ results are shown in Tab.\ \ref{tab_noisy}, while the mean of the noisy conditions and also the clean conditions for both codecs are listed. 
For the G.711-based narrowband experiments with noisy speech, both the postfilter~\cite{ITUTG711tollbox} and the proposed postprocessors hardly have an influence on the coded speech, with MOS-LQO differences being less than 0.04, and two insignificant degradations of only 0.01 MOS points being observed. On average, the postfilter and the proposed postprocessors have a MOS-LQO improvement in the range $0...0.02$ points.
For AMR-WB in noisy conditions, the cepstral domain postprocessors can improve or maintain the speech quality for most of the cases, with two exceptions: cafeteria noise (0.01 MOS points decrease) and road noise (0.03 MOS points decrease) both at 15 dB SNR. For car noise at both 15 and 20 dB obviously a speech quality improvement has been observed: 0.13 and 0.16 MOS points for structure \uppercase\expandafter{\romannumeral3}, 0.17 and 0.24 MOS points for structure \uppercase\expandafter{\romannumeral6}. The means over the noisy conditions show a MOS-LQO improvement of 0.05 points for structure \uppercase\expandafter{\romannumeral3} and 0.10 points for structure \uppercase\expandafter{\romannumeral6}. In summary and on average, \emph{both the G.711 postfilter and our proposed postprocessors do neither significantly improve nor distort noisy speech quality at the receiver}.

\textcolor{black}{Finally, in order to increase the robustness of the approach, we also trained the structure \uppercase\expandafter{\romannumeral3} model jointly with clean and noisy speech data. Four noise types\footnote{\textcolor{black}{The four noise types are: HOME-KITCHEN, HOME-LIVINGB, REVERB-POOL, and REVERB-CARPARK.}} from the QUT-NOISE database~\cite{dean2010qut} (noise types are different to the test data) are used to generate the 20 dB noisy training data, with the amount of the noisy data being one quarter of the clean data. As can be seen in Tab.~\ref{tab_noisy}, the model trained with noisy data (in the grey-shaded rows) achieves best performance on average and over the noisy conditions for both G.711 and AMR-WB. A test on clean data expectedly shows some reduced performance improvement. In summary and on average, \emph{the proposed postprocessor trained with additional noisy data can provide even some improvements in noisy conditions}.
	}

\begin{table}[]
	\begin{center}
		\footnotesize
		\begin{tabular}{m{0.001cm}<{\centering} m{2cm}<{\centering} | m{1.25cm}<{\centering} m{0.3cm}<{\centering} m{0.3cm}<{\centering} m{0.3cm}<{\centering} m{1.3cm}<{\centering}  m{0.001cm}}
			\cline{1-7}
			
			& & Frames per second & $L$ & $N$ & $F$ & MIPS & \\[18pt]
			\cline{1-7}
			& Time Domain& 100 &  80 & 15 & 55 & $3820\,(!)$ &\\[4pt]
			\cline{1-7}
			& \uppercase\expandafter{\romannumeral1}& 100 &  32 & 6 & 22 & $98.4$ &\\[4pt]
			\cline{3-7}
			\multirow{2}{*}{\rotatebox{90}{Cepstral Domain}} & \uppercase\expandafter{\romannumeral2}& 200 &  16 & 3 & 11 & $12.4$ &\\[4pt]
			\cline{3-7}
			& \uppercase\expandafter{\romannumeral3}& 100 &  32 & 6 & 22 & $98.4$ &\\[4pt]
			\cline{3-7}
			& \uppercase\expandafter{\romannumeral4}& 50 & 32 & 6 & 22 & $49.2$ &\\[4pt]
			\cline{3-7}
			& \uppercase\expandafter{\romannumeral5}& 50 & 32 & 6 & 22 & $49.2$ &\\[4pt]
			\cline{3-7}
			& \uppercase\expandafter{\romannumeral6}& 62.5 & 32 & 6 & 22 & $61.5$ &\\[4pt]
			\cline{1-7}
		\end{tabular}
	\end{center}
	\caption{\textbf{Computational complexity} im MIPS for the dominant convolutional operations in the CNN of each proposed framework structure in narrowband. The number of frames per second and the parameters of the CNN for all the proposed framework structures ($L$, $N$, and $F$) are also listed.}
	\label{tab_relative_complex}
	\vspace{-6pt}
\end{table}

\vspace{-4pt}
\subsection{Subjective Experiment} 
\label{subsec_subjective_exp}
In our CCR subjective listening test, 2 female and 12 male listeners participated, who are native German speakers stating to have no hearing impairment. An amount of 16 utterances from 4 speakers (2 female and 2 male) of the NTT speech database in German are subject to four test conditions following the processing plan in clean condition of Fig.\ \ref{figures_process_plan2}: The first is the \emph{direct} condition, resulting in the reference speech. The second is the \emph{legacy G.711} condition, providing speech transcoded by the G.711 codec. The third is the \emph{postfilter} condition, where G.711-transcoded speech has been enhanced by the ITU-T postfilter~\cite{ITUTG711tollbox}. The fourth is the \emph{proposed postprocessor} condition, where G.711-transcoded speech has been enhanced by our proposed postprocessor of structure \uppercase\expandafter{\romannumeral3} in the cepstral domain. Finally, all speech signals are converted to 48 kHz sampling rate. These four conditions result in six comparison cases in the subjective listening test (cf.\ Tab.  \ref{tab_subject}). 

In a preliminary informal subjective listening test we observed that an ideally very low 0-th cepstral coefficient turns out to assume slightly higher values after the CNN estimation, resulting in somewhat noisy speech pauses. Therefore, for the subjective listening test, we very slightly manipulate the CNN output as follows\textcolor{black}{\footnote{\textcolor{black}{
		Note that this manipulation naturally also degrades the instrumental values as given in Section~\ref{clean_conditions}. For structure \uppercase\expandafter{\romannumeral3} in Tab.~\ref{tab_dif_codecs}, e.g., we observed deviations in the range $[-0.12\,...\!+\!\!0.02]$ over languages and codecs, however, still exceeding all legacy codecs and the postfilter~\cite{ITUTG711tollbox} in instrumental metrics.
		}}}
		
\begin{equation}
	\label{equ_C0_manip}
	\hat{c}_{\text{env}}(\ell,0)\rightarrow 
	\begin{cases}
		\hat{c}_{\text{env}}(\ell,0), ~\text{if} ~ \hat{c}_{\text{env}}(\ell,0)\geqslant C_{0}\\
		\hat{c}_{\text{env}}(\ell,0)-\gamma_{0}, ~\text{else},
	\end{cases}
\end{equation}
with $C_{0}\!=\!-1650$ and $\gamma_{0}\!=\!1000$.

The participants of the subjective listening test rated the speech using an \texttt{AKG K-271 MKII} headphone from a computer with external \texttt{RME Fireface 400} sound card. The participants were equally assigned to one of two disjoint sets, where the speech is balanced over the comparison cases and the speakers. 
Each participant familiarized himself with all the comparison cases and was asked to choose a proper volume on the basis of 12 sample pairs in the familiarization phase. Then, each participant evaluated 72 sample pairs in the main test phase, where 36 sample pairs are presented in both sample orders. 

\begin{figure}[tp]
	
	\psfrag{100}[tc][tc]{\footnotesize $100$}
	\psfrag{200}[tc][tc]{\footnotesize $200$}
	\psfrag{300}[tc][tc]{\footnotesize $300$}
	\psfrag{400}[tc][tc]{\footnotesize $400$}
	\psfrag{500}[tc][tc]{\footnotesize $500$}
	\psfrag{600}[tc][tc]{\footnotesize $600$}
	\psfrag{700}[tc][tc]{\footnotesize $700$}
	
	\psfrag{0}[cr][cr]{\footnotesize $0$}
	\psfrag{64}[cr][cr]{\footnotesize $64$}
	\psfrag{128}[cr][cr]{\footnotesize $128$}
	\psfrag{192}[cr][cr]{\footnotesize $192$}
	\psfrag{256}[cr][cr]{\footnotesize $256$}
	
	\psfrag{Frame}[tc][bc]{\footnotesize Frame $\ell$}
	\psfrag{FrequencyBins}[bc][tc]{\footnotesize Frequency bins $k$}
	
	\psfrag{ReferenceSpeech}[cc][cc]{\footnotesize  }
	\psfrag{CodedSpeech}[cc][cc]{\footnotesize  }
	\psfrag{PostprocessedSpeech}[cc][cc]{\footnotesize  }

	\centering
	\includegraphics[height=\columnwidth,angle=270]{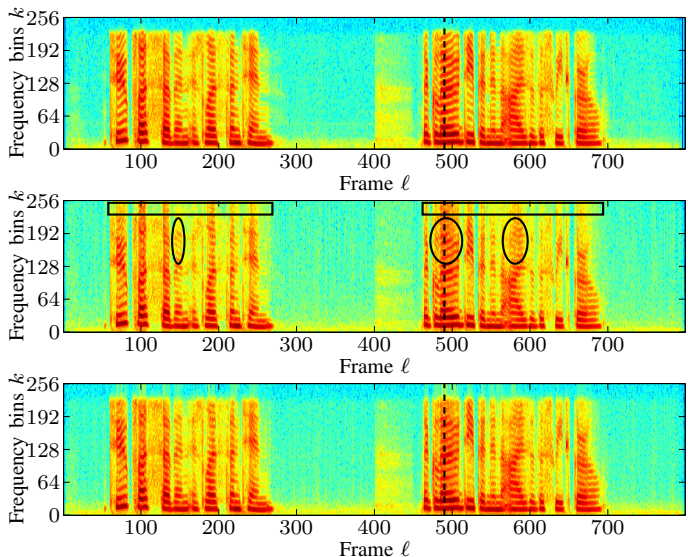}   
	\textcolor{black}{\caption{Narrowband spectrograms of an utterance: reference speech (top), G.726-coded speech (center), and postprocessed speech (bottom). Characteristic time-frequency regions and frame $\ell\!=\!490$ are marked.}}
	\label{fig_specgram1}
\end{figure}

In Tab.\ \ref{tab_subject}, the CMOS and respective 95\% confidence interval ($CI_{95}$) for the six CCR comparison cases are shown. \emph{All results turned out to be significant}.
We can see a clear 1.76 CMOS points advantage for the comparison of legacy G.711 vs. direct. 
For the cases where the direct condition is compared to the postfilter~\cite{ITUTG711tollbox} and the proposed postprocessor of structure \uppercase\expandafter{\romannumeral3} in the cepstral domain, 0.28 and -0.18 CMOS points are obtained, respectively. This means that the speech enhanced by the proposed postprocessor is more similar to the uncoded speech (in direct condition), and even slightly but {\it significantly preferred to uncoded speech}. To the best knowledge of the authors, such a result has never been reported before. For details, however, see~\cite{zhao2018enhancement_g711_ITG_submit}. Our only explanation is the very low-energy in speech pause during the direct condition, which, of course, we are not allowed to manipulate. 
Relative to the legacy G.711 condition, the ITU-T postfilter~\cite{ITUTG711tollbox} already shows a significant 1.45 CMOS points advantage, while the proposed postprocessor performs even better, obtaining 1.77 CMOS points above the legacy G.711 condition. When the proposed postprocessor is directly compared to the ITU-T postfilter~\cite{ITUTG711tollbox}, a better performance of 0.36 CMOS points is obtained. Finally, we conclude that the proposed postprocessor improves the quality of G.711-coded speech more effectively as the ITU-T postfilter~\cite{ITUTG711tollbox} does.

\begin{figure}[tp]
	
	\psfrag{100}[cr][cr]{\footnotesize $100$}
	\psfrag{50}[cr][cr]{\footnotesize $50$}
	\psfrag{80}[cr][cr]{\footnotesize $80$}
	\psfrag{60}[cr][cr]{\footnotesize $60$}
	\psfrag{40}[cr][cr]{\footnotesize $40$}
	\psfrag{20}[cr][cr]{\footnotesize $20$}
	\psfrag{0}[cr][cr]{\footnotesize $0$}
	\psfrag{-20}[cr][cr]{\footnotesize $-20$}
	\psfrag{01}[tl][tr]{\footnotesize $0$}
	\psfrag{02}[cl][br]{\footnotesize $0$}
	
	\psfrag{300}[cr][cr]{\footnotesize $300$}
	\psfrag{200}[cr][cr]{\footnotesize $200$}
	\psfrag{-100}[cr][cr]{\footnotesize $-100$}
	\psfrag{-200}[cr][cr]{\footnotesize $-200$}
	
	\psfrag{32}[tc][tc]{\footnotesize $32$}
	\psfrag{64}[tc][tc]{\footnotesize $64$}
	\psfrag{96}[tc][tc]{\footnotesize $96$}
	\psfrag{128}[tc][tc]{\footnotesize $128$}
	\psfrag{160}[tc][tc]{\footnotesize $160$}
	\psfrag{192}[tc][tc]{\footnotesize $192$}
	\psfrag{224}[tc][tc]{\footnotesize $224$}
	\psfrag{256}[tc][tc]{\footnotesize $256$}
	
	\psfrag{2}[tc][tc]{\footnotesize $2$}
	\psfrag{4}[tc][tc]{\footnotesize $4$}
	\psfrag{10}[tc][tc]{\footnotesize $10$}
	\psfrag{15}[tc][tc]{\footnotesize $15$}
	\psfrag{201}[tc][tc]{\footnotesize $20$}
	\psfrag{251}[tc][tc]{\footnotesize $24$}
	\psfrag{30}[tc][tc]{\footnotesize $30$}
	
	\psfrag{Amplitute}[bc][tc]{\footnotesize $10\text{log}(|S|^2)$}
	\psfrag{FrequencyBins}[tc][bc]{\footnotesize Frequency bins $k$}
	\psfrag{AmplituteCeps}[bc][tc]{\footnotesize Cepstral coefficients}
	\psfrag{CepstrumBins}[tc][bc]{\footnotesize Cepstral coefficient indices $m$}
	
	\psfrag{ReferenceSpeeeech}[cc][cc]{\scriptsize Reference Speech}
	\psfrag{CodedSpeeeech}[cl][cl]{\scriptsize G.726-Coded Speech}
	\psfrag{PostprocessedSpeeeech}[cc][cc]{\scriptsize Postprocessed Speech}
	
	\psfrag{title}[cc][cc]{\footnotesize}
	
	\centering
	\hspace*{11pt}\includegraphics[width=0.96\columnwidth]{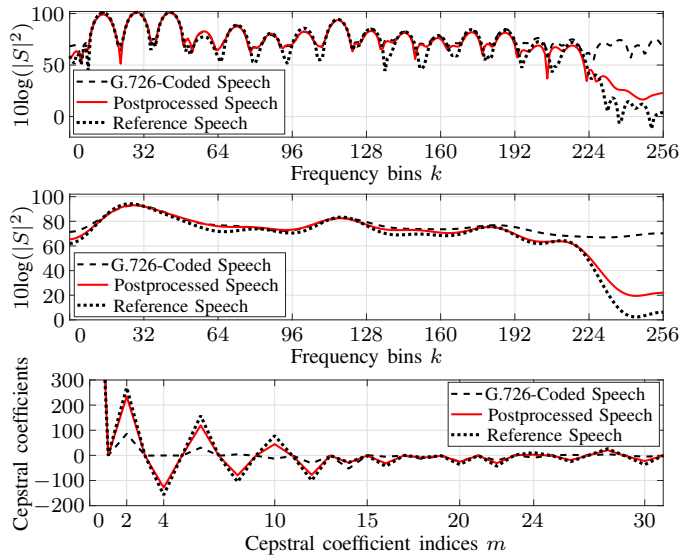}   
	\vspace{-11pt}
	\textcolor{black}{\caption{Amplitude spectrum (top), spectral envelope (center), both on a logarithmic scale, and DCT-\uppercase\expandafter{\romannumeral2} type of cepstral coefficients (bottom) for frame $\ell\!=\!490$ (see Fig.\ 11) of the narrowband reference speech, G.726-coded speech, and postprocessed speech, respectively. }}
	\label{fig_spec_env}
	\vspace{-10pt}
\end{figure}

\vspace{-4pt}
\subsection{Complexity Analysis}
\label{subsec_relative_complex}
The complexity of the time domain approach basically lies in the computations for the CNN. Neglecting the operations in Fig.\ \ref{figures_CNN_topology} of max pooling, upsampling and skip connection addition, the complexity-dominant convolutional operations of the CNN amount to about $10.5\!\cdot\!NLF^2\!+\!2\!\cdot\!NLF$ multiply/accumulates (MACs) per frame of the time domain approach, with $L$ being the frame length (i.e., 10 ms of speech samples) and $N$, $F$ being the parameters of CNN. 
For the cepstral domain approaches, the number of MACs in the CNN follows the same expression as the time domain approach, with $L\!=\!\left | \mathcal{M}_{\text{env}} \right |\!=\!6.25\%\cdot K$. Moreover, some operations are required besides the computations in the CNN: FFT, IFFT, DCT-II, and IDCT-II all have a computational complexity of $\text{O}(K \text{log} K)$~\cite{cooley1965algorithm,narasimha1978computation}. 

In order to show the complexity of the proposed CNN-based postprocessors, the million instructions ($=$ MACs) per second (MIPS) for the convolutional operations in the CNN of each proposed framework structure in narrowband are shown in Tab.\ \ref{tab_relative_complex}. Note that the values of $L$, $N$, and $F$ are doubled in wideband, resulting in a larger number of MACs per second compared to that in narrowband. We see that the time domain postprocessor requires a lot of computations, while the cepstral domain postprocessors have moderate complexity in terms of MIPS, roughly in the order of magnitude of a modern speech codec. As an outlook to future work, however, it might be attractive to reduce the complexity of the models further by methods such as teacher-student learning.

\vspace{-3pt}
\section{Conclusions} \label{sec_conclusion}
In this work, we propose two different CNN-based postprocessing approaches in the time domain and the cepstral domain, including six different framework structures for the latter, to enhance coded speech in a system-compatible manner. 
The proposed postprocessors in both domains are evaluated for various narrowband and wideband speech codecs in clean, tandeming, error-prone transmission and noisy conditions, and they are compared to an ITU-T postfilter~\cite{ITUTG711tollbox} as postprocessing baseline for G.711. 
The proposed postprocessor improves speech quality in terms of PESQ by up to 0.25 MOS-LQO points for G.711, 0.30 points for G.726, 0.82 points for G.722, and 0.26 points for AMR-WB.
In a subjective CCR listening test, the proposed postprocessor on G.711-coded speech exceeds the speech quality of an ITU-T-standardized postfilter by 0.36 CMOS points, and obtains a clear preference of 1.77 CMOS points compared to G.711, even significantly exceeding the quality of uncoded speech. The source code for the cepstral domain approach to enhance G.711-coded speech is available at \href{https://github.com/ifnspaml/Enhancement-Coded-Speech}{https://github.com/ifnspaml/Enhancement-Coded-Speech}.


%



\begin{figure}[tp]
	
	\psfrag{100}[tc][tc]{\footnotesize $100$}
	\psfrag{200}[tc][tc]{\footnotesize $200$}
	\psfrag{300}[tc][tc]{\footnotesize $300$}
	\psfrag{400}[tc][tc]{\footnotesize $400$}
	\psfrag{500}[tc][tc]{\footnotesize $500$}
	\psfrag{600}[tc][tc]{\footnotesize $600$}
	\psfrag{700}[tc][tc]{\footnotesize $700$}
	
	\psfrag{0}[cr][cr]{\footnotesize $0$}
	\psfrag{64}[cr][cr]{\footnotesize }
	\psfrag{128}[cr][cr]{\footnotesize $128$}
	\psfrag{192}[cr][cr]{\footnotesize }
	\psfrag{256}[cr][cr]{\footnotesize $256$}
	\psfrag{320}[cr][cr]{\footnotesize }
	\psfrag{384}[cr][cr]{\footnotesize $384$}
	\psfrag{448}[cr][cr]{\footnotesize }
	\psfrag{512}[cr][cr]{\footnotesize $512$}
	
	\psfrag{Frame}[tc][bc]{\footnotesize Frame $\ell$}
	\psfrag{FrequencyBins}[bc][tc]{\footnotesize Frequency bins $k$}
		
	\centering
	\includegraphics[height=\columnwidth,angle=270]{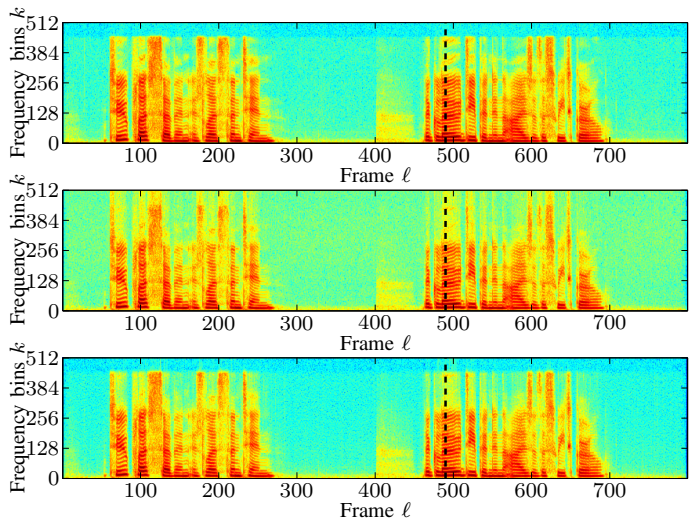}   
	\textcolor{black}{\caption{Wideband spectrograms of an utterance: reference speech (top), G.722-coded speech (center), and postprocessed speech (bottom). Frame $\ell\!=\!490$ is marked.}}
	\label{fig_specgram_wb}
	\vspace{-5pt}
\end{figure}

\vspace{-10pt}
\textcolor{black}{\appendix} 
In this Appendix we will provide some further detailed analysis of our postprocessor in certain conditions.

 We take the speech file \texttt{am02f065} from the NTT speech database in American English as an example, and plot the spectrograms for preprocessed speech (i.e., reference speech, see Fig.~\ref{figures_process_plan2}), G.726-coded speech, and enhanced speech by the cepstral domain approach with structure \uppercase\expandafter{\romannumeral3} in Fig.\;11. The spectral analysis settings are identical to the framework structure \uppercase\expandafter{\romannumeral3} (see Tab.\ \ref{table_structures}). Comparing top and center subplots, the G.726 coding adds signal contents to the high frequencies (marked by rectangles) \textcolor{black}{and distorts/weakens spectral envelope (marked by ovals)}. For the enhanced speech in the bottom, the high frequency coding noise is effectively eliminated, and the \textcolor{black}{spectral envelope is somewhat being} restored and enhanced towards the reference speech \textcolor{black}{spectral envelope}. 

\textcolor{black}{In order to show that the improvement of the postprocessor is not only based on a trivial postfilter simply suppressing frequencies beyond 3.5 kHz, we did a brief PESQ MOS measurement of coded speech with high frequencies simply removed: A lowpass filter cutting off at 3.5 kHz is applied to the coded speech. The FLAT filter is used here, along with the up- and downsampling, since the FLAT filter works at 16 kHz. It turns out that the PESQ MOS scores did not even change after this trivial postfiltering for both narrowband codecs (G.711 and G.726) for both languages. This maybe surprising result shows that the major speech quality improvement does not at all come from the trivial filtering, but supports our proposed postprocessor which also acts on lower frequencies.}

\begin{figure}[tp]
	
	\psfrag{100}[cr][cr]{\footnotesize $100$}
	\psfrag{50}[cr][cr]{\footnotesize $50$}
	\psfrag{80}[cr][cr]{\footnotesize $80$}
	\psfrag{60}[cr][cr]{\footnotesize $60$}
	\psfrag{40}[cr][cr]{\footnotesize $40$}
	\psfrag{20}[cr][cr]{\footnotesize $20$}
	\psfrag{0}[cr][cr]{\footnotesize $0$}
	\psfrag{-20}[cr][cr]{\footnotesize $-20$}
	\psfrag{01}[tl][tr]{\footnotesize $0$}
	\psfrag{02}[cl][br]{\footnotesize $0$}
	
	\psfrag{300}[cr][cr]{\footnotesize $300$}
	\psfrag{200}[cr][cr]{\footnotesize $200$}
	\psfrag{-100}[cr][cr]{\footnotesize $-100$}
	\psfrag{-200}[cr][cr]{\footnotesize $-200$}
	
	\psfrag{32}[tc][tc]{\footnotesize $32$}
	\psfrag{64}[tc][tc]{\footnotesize $64$}
	\psfrag{96}[tc][tc]{\footnotesize $96$}
	\psfrag{128}[tc][tc]{\footnotesize $128$}
	\psfrag{160}[tc][tc]{\footnotesize $160$}
	\psfrag{192}[tc][tc]{\footnotesize $192$}
	\psfrag{224}[tc][tc]{\footnotesize $224$}
	\psfrag{256}[tc][tc]{\footnotesize $256$}
	\psfrag{320}[tc][tc]{\footnotesize $320$}
	\psfrag{384}[tc][tc]{\footnotesize $384$}
	\psfrag{448}[tc][tc]{\footnotesize $448$}
	\psfrag{512}[tc][tc]{\footnotesize $512$}
	
	\psfrag{51}[tc][tc]{\footnotesize $5$}
	\psfrag{201}[tc][tc]{\footnotesize $20$}
	\psfrag{401}[tc][tc]{\footnotesize $40$}
	\psfrag{501}[tc][tc]{\footnotesize $50$}
	\psfrag{601}[tc][tc]{\footnotesize $60$}
	
	\psfrag{10}[tc][tc]{\footnotesize $10$}
	\psfrag{15}[tc][tc]{\footnotesize $15$}
	\psfrag{201}[tc][tc]{\footnotesize $20$}
	\psfrag{251}[tc][tc]{\footnotesize $25$}
	\psfrag{30}[tc][tc]{\footnotesize $30$}

	\psfrag{25}[tc][tc]{\footnotesize $25$}
	\psfrag{35}[tc][tc]{\footnotesize $35$}
	\psfrag{45}[tc][tc]{\footnotesize $45$}
	\psfrag{55}[tc][tc]{\footnotesize $55$}
	
	\psfrag{Amplitute}[bc][tc]{\footnotesize $10\text{log}(|S|^2)$}
	\psfrag{FrequencyBins}[tc][bc]{\footnotesize Frequency bins $k$}
	\psfrag{AmplituteCeps}[bc][tc]{\footnotesize Cepstral coefficients}
	\psfrag{CepstrumBins}[tc][bc]{\footnotesize Cepstral coefficient indices $m$}
	
	\psfrag{ReferenceSpeeeech}[cc][cc]{\scriptsize Reference Speech}
	\psfrag{CodedSpeeeech}[cl][cl]{\scriptsize G.722-Coded Speech}
	\psfrag{PostprocessedSpeeeeech}[cc][cc]{\scriptsize Postprocessed Speech}
	
	\psfrag{title}[cc][cc]{\footnotesize}
	
	\centering
	\vspace*{3pt}
	\hspace*{11pt}\includegraphics[width=0.96\columnwidth]{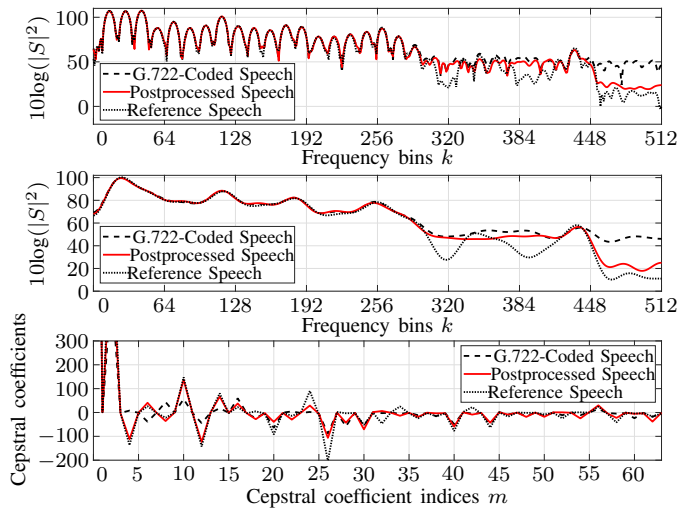}   
	\vspace{-11pt}
	\textcolor{black}{\caption{Amplitude spectrum (top), spectral envelope (center), both on a logarithmic scale, and DCT-\uppercase\expandafter{\romannumeral2} type of cepstral coefficients (bottom) for frame $\ell\!=\!490$ (see Fig.\ 13) of the wideband reference speech, G.722-coded speech, and postprocessed speech, respectively. }}
	\label{fig_spec_env_wb}
	\vspace{-10pt}
\end{figure}

\textcolor{black}{
	As the proposed cepstral domain approach intends to improve the spectral envelope, we zoom into the frame $\ell\!=\!490$ (dashed line in Fig.\ 11) to have a clear view of the spectral envelope. In Fig.\ 12, the logarithmic spectrum $10\text{log}(|S(k)|^2)$ of the selected frame is drawn in the top, and of the spectral envelope in the middle, obtained by keeping the first 32 cepstral coefficients and setting the other cepstral coefficients to zero, i.e., lowpass liftering. As we can see, the spectral envelope of the enhanced speech is closer to the reference speech as is the coded speech. This holds particularly for higher frequencies, which shows the efficacy of the proposed cepstral domain approach. Finally, we take a look on the cepstral coefficients in the bottom of Fig.\ 12. It can be seen that the cepstrum of the postprocessed speech is also closer to the reference speech, enhancing the cepstrum of coded speech not only into the same direction (e.g., $m\!=\!2$ and $m\!=\!4$), but even reversing the sign to better approach the reference cepstrum (e.g., $m\!=\!10$ and $m\!=\!24$).
}

\textcolor{black}{
	We also conduct the same analysis as we did for G.726 above for G.722-coded speech (wideband speech) and the results are presented in Fig.\ 13 and Fig.\ 14. It can be seen that a similar trend holds: The spectral envelope and particularly the cepstral coefficients of the postprocessed speech are closer to those of the reference speech. 
}

In order to explain the reason of the impressive MOS score improvement of G.722-coded speech (Tab.\ \uppercase\expandafter{\romannumeral4}) after the proposed postprocessing, we also conducted a  similar experiment as for G.726 above, to identify the improvement of using a simple P.341 filter after G.722-coded speech cutting off at around 7 kHz for wideband speech. It turns out that the improvements of this simple filtering are already non-negligible, which are 0.32 and 0.31 PESQ MOS for American English and German, respectively\textcolor{black}{\footnote{\textcolor{black}{This $\sim$0.3 MOS score improvement by simple P.341 filtering is due to the reference signal we used, which is also a P.341 filter output with similar frequency content. Accordingly, from an informal subjective listening, there is no big difference between G.722-coded speech and G.722-coded speech followed by this simple P.341 filter, meaning that P.341 postfiltering in our setup is no practically valid postprocessor. }}}. This result only partly explains the reason of the improvement, since the postprocessed speech shows a similar cutting-off effect at around 7 kHz (see Fig.\ 13). The further improvement of the postprocessing in the cepstral domain being roughly 0.5 MOS points now can be dedicated to the enhanced spectral envelope. 


\vspace{9pt}
\section*{Acknowledgment}
The authors would like to thank S. Elshamy for providing an implementation of both DCT-\uppercase\expandafter{\romannumeral2} and the IDCT-\uppercase\expandafter{\romannumeral2} and J. Abel for advice concerning the setup of the subjective listening test. 

\ifCLASSOPTIONcaptionsoff
  \newpage
\fi



%

\bibliographystyle{IEEEtran}
\bibliography{./IEEEabrv,./thesis_zhao}

\vfill


\end{psfrags}
\end{document}